\documentclass[preprint]{aastex}
\usepackage[normalem]{ulem}

\shorttitle{A New Formation Scenario for NGC\,2419}
\shortauthors{Br\"uns et al.}

\begin{document}

\title{A New Formation Scenario for the Milky Way Cluster NGC\,2419}

\author{R.C. Br\"uns and P. Kroupa}
\affil{Argelander-Institut f\"ur Astronomie, Universit\"at Bonn, Auf dem H\"ugel 71, D-53121 Bonn, Germany}
\email{rcbruens@astro.uni-bonn.de, pavel@astro.uni-bonn.de}

\begin{abstract} \label{abstract}
We present a new formation scenario for NGC\,2419, which is one of the most luminous,
one of the most distant, and as well one of the most extended globular clusters of the Milky Way.
We propose that NGC\,2419 is the remnant of a merged star cluster complex, which was possibly 
formed during an interaction between a gas-rich galaxy and the Milky Way.  
To test this hypothesis, we performed numerical simulations of 27 different 
models of star cluster complexes (CCs) moving on a highly eccentric orbit in the Galactic halo. 
We vary the CC mass, the CC size, and the initial distribution of star clusters in the CC to 
analyze the influence of these parameters on the resulting objects.

In all cases, the vast majority of star clusters merged into a stable object. The derived 
parameters mass, absolute V-band magnitude, effective radius, velocity dispersion and the 
surface brightness profile are, for a number of models, in good agreement with those 
observed for NGC\,2419. Despite the large range of CC sizes, the effective radii
of the merger objects are constrained to a relatively small interval. A turnover in the 
$r_{\rm eff}$ vs. $M_{\rm encl}$ space leads to degenerate states, i.e. relatively compact 
CCs can produce an object with the same structural parameters as a more massive and larger CC. 
In consequence, a range of initial conditions can form a merger object comparable to NGC\,2419 
preventing us to pinpoint the exact parameters of the original CC, which formed NGC\,2419.

We conclude that NGC\,2419 can be well explained by the merged cluster complex scenario.
Some of the recently discovered stellar streams in the Galactic halo might be related to 
the parent galaxy, which produced the cluster complex in our scenario. Measurements 
of the proper motion of NGC\,2419 are necessary to prove an association with one of the stellar 
streams.
\end{abstract}

\keywords{Galaxy: evolution --- globular clusters: individual (NGC2419) --- methods: numerical} 

\section{INTRODUCTION} \label{introduction}
Globular clusters (GCs) are among the oldest radiant objects in the universe. They
have typically compact sizes with half-light radii of a few pc and absolute V-band magnitudes 
between $M_{\rm V} = -5$ to $M_{\rm V} = -10$ corresponding roughly to masses between 
$10^4\,M_\odot$ and $10^6\,M_\odot$ \citep[see][and references therein]{brodie06}.

\begin{figure}
\epsscale{1.0}
\plotone{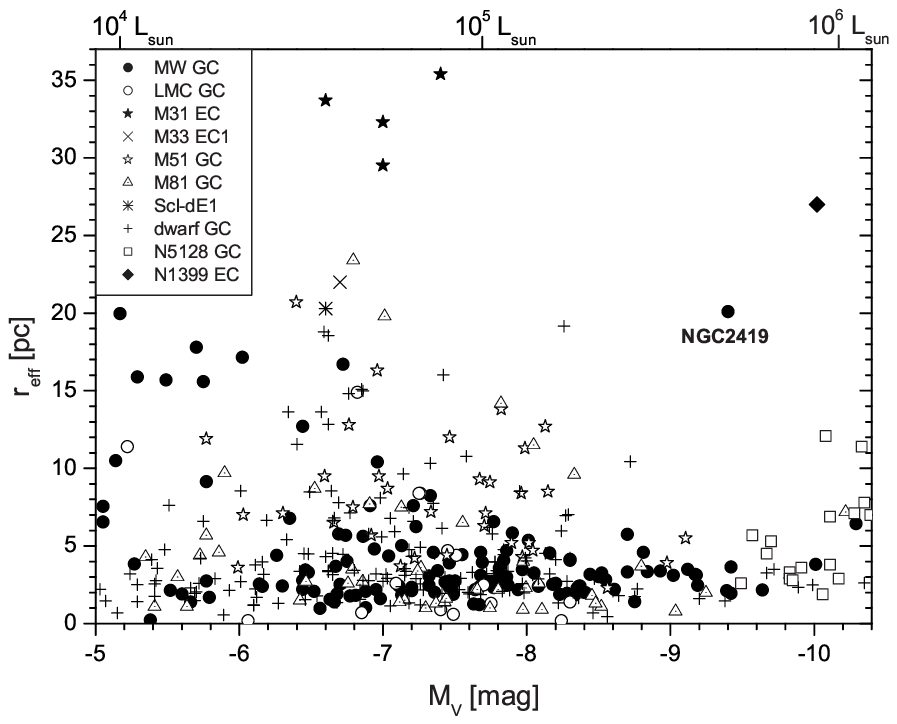}
\caption{Overview of GCs of the Milky Way \citep{harris}, the LMC \citep{mackey04,vandenbergh04}, 
M51, and M81 \citep{chandar04}, NGC\,5128 \citep{harris02}, 68 dwarf galaxies \citep{georgiev} 
and the ECs from M31 \citep{huxor04,mackey06}, M33 \citep{stonkute}, Scl-dE1 \citep{dacosta}, 
and one EC from NGC\,1399 \citep{richtler} in the $r_{\rm eff}$ vs. $M_{\rm V}$ space. NGC\,2419 
is in a rather isolated position. \label{fig1}}
\end{figure}

The Milky Way has a rich GC system containing 150 GCs \citep{harris}. Only 13 GCs (or 9\%) have an 
effective radius larger than 10~pc (see Figure \ref{fig1}). Most of these extended clusters (ECs) are 
fainter than about $M_{\rm V} = -7$, only NGC\,2419 has a high luminosity of about $M_{\rm V} = -9.4$ mag. 
While all GCs with a comparable luminosity are quite compact, NGC\,2419 has a projected half-light or
effective radius of about 20~pc. The structural properties of NGC\,2419 derived by various 
authors are displayed in Table \ref{tbl-1}. NGC\,2419, located at ($l$,$b$)=(180\fdg4, 25\fdg2), 
is one of the most metal-poor GCs with a metallicity of \mbox{[Fe/H] = --2.12} \citep{harris} at 
a Galactocentric distance of about 91.5 kpc and a heliocentric distance of 
82.4 kpc \citep{harris}. 
NGC\,2419 shows a single stellar population with an age of 12 to 13 Gyr \citep{salaris}. 

Figure \ref{fig1} also shows the GCs from the LMC \citep{mackey04,vandenbergh04}, 
M51, and M81 \citep{chandar04}, from 68 dwarf galaxies of nearby loose galaxy groups \citep{georgiev}, 
and the nearby elliptical galaxy NGC\,5128 \citep{harris02}, as well as the ECs from M31 \citep{huxor04,mackey06}, 
M33 \citep{stonkute}, Scl-dE1 \citep{dacosta}, and EC 90:12 from NGC\,1399
\citep{richtler}. It is evident that NGC\,2419 has a rather isolated location 
in the $M_{\rm V}$ vs. $r_{\rm eff}$ space. 

The unusual size and mass of NGC\,2419 led to the speculation that NGC\,2419 is not a GC, but 
the stripped core of a dSph galaxy captured by the Milky Way \citep[e.g.][]{vandenbergh04}. 
A stripped core should, however, show multiple stellar populations from the progenitor galaxy. 
In contrast, \cite{ripepi} demonstrated that NGC\,2419 shows no sign of multiple stellar populations
and concluded that NGC\,2419 is not a stripped core of a dwarf galaxy.
\cite{ripepi} also found that NGC\,2419 is an Oosterhoff II cluster. They concluded that an 
extra-galactic origin of NGC\,2419 is unlikely as the GCs of the Galactic dwarf galaxies  
usually show parameters between Oosterhoff types I and II. On the other hand, differences between 
NGC\,2419 and typical values of GCs associated with Galactic dwarf spheroidal galaxies do not rule 
out an extra-galactic origin of NGC\,2419 per se. For instance, five out of twelve GCs from the LMC are 
Oosterhoff II clusters \citep{catelan09}.

High-resolution imaging of gas-rich galaxies experiencing major interactions using the Hubble-Space
Telescope has resolved regions with very intense star formation bursts. The bursting regions 
are typically located within the severely perturbed disks or tidal tails and are constrained to
small complexes that contain a few to hundreds
of young massive star clusters. Examples of such systems are
the knots in the Antennae galaxies \citep{whitmore95, whitmore99}, 
the complexes in the NGC\,7673 star-burst \citep{homeier}, M82 \citep{konstantopoulos}, 
Arp24 \citep{cao}, the ``bird's head galaxy'' NGC\,6745 \citep{de_grijs03}, 
NGC\,6946 \citep{larsen02}, Stephan's Quintet \citep{gallagher01}, and NGC\,922 
\citep{pellerin}.

The masses of such complexes vary from $10^6\,$M$_\odot$ up to a few~$10^8\,$M$_\odot$.
\cite{bastian06} observed star cluster complexes (CCs) in the Antennae with masses of the order 
$\approx 10^{6}$~M$_{\odot}$ and sizes of the order 100 to 200 pc. \cite{pellerin} found 
young massive CCs with masses between $10^6$ M$_\odot$ and $10^{7.5}$ M$_\odot$ 
and diameters between 600 pc and 1200 pc in the collisional ring galaxy NGC\,922.

\cite{mengel08} observed individual young ($\approx$10 Myr) star clusters associated with 
cluster complexes in the Antennae and NGC\,1487. They compared dynamical mass estimates with 
derived photometric masses and found them in excellent agreement, implying that most of them 
survived the gas removal phase and are bound stellar objects. These young clusters 
are sufficiently stable to be used as building blocks for numerical simulations.
\cite{bastian09} found three 200 to 500 Myr old, apparently stable clusters in the Antennae with 
very high radial velocities relative to the galactic disk, indicating that these star clusters 
will most likely become future halo objects. One cluster is surrounded by so far unmerged stellar 
features in its vicinity.

It has already been shown that CCs can merge to form a variety of spheroidal stellar-dynamical 
objects, such as ultra-compact dwarf galaxies (UCDs), faint fuzzies and possibly dwarf 
spheroidal galaxies \citep{krou98,fellhauer02a, fellhauer02b,bruens09}. In particular, the 
young UCD W3 is most naturally understood to be a merged massive CC \citep{fellhauer05}. 

In the present paper, we apply these ideas to NGC\,2419 by analyzing how the structural parameters 
of the final merger objects in the outer Galactic halo correlate with the underlying CC parameters and 
compare them with the observed parameters of NGC\,2419.

In Section \ref{simulations}, we describe the method and the parameters used for the calculations. 
Section \ref{results} presents the results of the simulations, which will be discussed 
in Section \ref{discussion}. In Section \ref{summary} we provide a summary.

\newcommand{\cpp}{C{\footnotesize\raise.4ex\hbox{+\kern-.2em+}}}
\newcommand{\sbpp}{{\scshape Superbox{\footnotesize\raise.4ex\hbox{+\kern-.2em+}}}}
\newcommand{\subo}{{\scshape Superbox}}

\section{NUMERICAL METHOD AND SET-UP} \label{simulations}
The formation scenario described in this paper starts with newly born complexes of star clusters
in the Galactic halo. We model the dynamical evolution of various CCs leading to merger objects. 
We do not, however, consider the galaxy-galaxy interaction, which formed the CCs in the first place.

\subsection{The Code \sbpp\ } \label{method}

\sbpp\ developed by \cite{metz} is a new implementation of the particle-mesh code \subo\ 
\citep{bien,fell00}. Differences in the implementation of the algorithm make the 
new code \sbpp\ much more efficient than the former. \subo\ had been implemented in the FORTRAN 
language with a particular focus on the minimization of usage of the random access memory (RAM), 
which is no longer a big issue in current days. \sbpp\ is now being implemented in the modern 
\cpp\ programming language using object oriented programming techniques. The algorithm has been 
implemented with a focus on the performance of the code, but at the same time keeping memory 
consumption at a low level. \sbpp\ makes particular optimal use of modern multi-core 
processor technologies.

The code solves the Poisson equation on a system of Cartesian grids.  
The local universe is covered by a fixed coarse grid which contains the orbit of the CC around 
the center of the Milky Way.  
In order to get good resolution of the star clusters two grids with high and medium resolution 
are focused on each star cluster following their trajectories. The individual high resolution 
grids cover an entire star cluster, whereas the medium resolution grids of every star cluster 
embed the whole initial CC. All grids 
contain 128$^{3}$ grid cells. The CC orbits in an analytical Galactic potential 
(see Sect. \ref{potential_mw}). For each particle in the CC the acceleration from the galactic 
potential is added as an analytical formula to the grid-based acceleration computed by solving 
the Poisson equation.
The coordinate system is chosen such that the disk of the 
\mbox{Milky Way} lies in the x-y-plane with origin at the Galactic center.

\subsection{Orbit of NGC\,2419} \label{orbit_ngc2419}
\subsubsection{Gravitational Potential of the Milky Way} \label{potential_mw}

In our computations the Milky Way is represented by an analytical potential, which consists of a 
\mbox{disk-,} a bulge-, and a halo component.

The disk is modeled by a Miyamoto-Nagai potential \citep{miya1975},
\begin{eqnarray}
\Phi_{\rm disk}(R,z)= -\frac{G\,M_{\rm d}}{\sqrt{R^{2}+(a_{\rm d}+\sqrt{z^{2}+b_{\rm d}^{2}})^{2}}},\label{pot_mw1} 
\end{eqnarray}
with $M_{\rm d}=1.0 \times 10^{11}$ M$_{\odot}$, $a_{\rm d}=6.5$ kpc, and $b_{\rm d}=0.26$ kpc. 
The bulge is represented by a Hernquist potential \citep{hern1990},
\begin{eqnarray}
\Phi_{\rm bulge}=-\frac{G \, M_{\rm b}}{r+a_{\rm b}},\label{pot_mw2} 
\end{eqnarray}
with $M_{\rm b}=3.4 \times 10^{10}$ M$_{\odot}$, and $a_{\rm b}=0.7$ kpc.
The halo is a logarithmic potential,
\begin{eqnarray}
\Phi_{\rm halo}(r)=\frac{1}{2} \, v_{\rm 0}^{2} \, {\rm ln}(r^{2}+r_{\rm halo}^{2}), \label{pot_mw3} 
\end{eqnarray}
with $v_{\rm 0}=186.0$ km s$^{-1}$, and $r_{\rm halo}=12.0$ kpc.\\
This set of parameters gives a realistic rotation curve for the Milky Way.

\subsubsection{Orbital Parameters} \label{orbit}
The calculation of an orbit for NGC\,2419 requires, next to the spatial coordinates,
a good knowledge of the actual velocity vector of the cluster. As no proper motion measurements for 
NGC\,2419 are available, the orbit cannot be properly fixed without major assumptions.

\cite{king62} suggested that the tidal radius of a GC is determined by the orbital position with the 
highest gravitational force, which is the perigalacticon. 
\cite{vandenbergh95} used this method to estimate a perigalactic distance of 20 kpc for NGC\,2419.
While this method provides only a very rough estimate of the perigalactic distance, we use it 
to narrow down the range of possible orbits. 

NGC\,2419 is one of the outermost Galactic GCs at a distance of $R_{\rm gal}$ = 91.5 kpc \citep{harris}. 
The very large distance and the observed low (heliocentric) negative radial velocity of 
$v_{\rm hel}$ = --20.3 km s$^{-1}$ \citep{baumgardt}, which corresponds to a Galactic-standard-of-rest
velocity of $v_{\rm GSR}$ = --26.8 km s$^{-1}$, suggests that NGC\,2419 is close to its apogalacticon. 

We choose a velocity vector for NGC\,2419 in a way that it 
\begin{itemize} 
\item is consistent with the observed radial velocity,
\item produces an orbit between Galactic radii 20 and about 92 kpc,
\item is neither a polar orbit nor an orbit within the Galactic plane.
\end{itemize}
The velocity vector ($v_{\rm x}$, $v_{\rm y}$, $v_{\rm z}$) = (45 km s$^{-1}$, -56 km s$^{-1}$, 47 km s$^{-1}$) will 
be used as the current velocity vector of NGC\,2419 for calculating the orbit. 

\begin{figure}
\epsscale{1.0}
\plotone{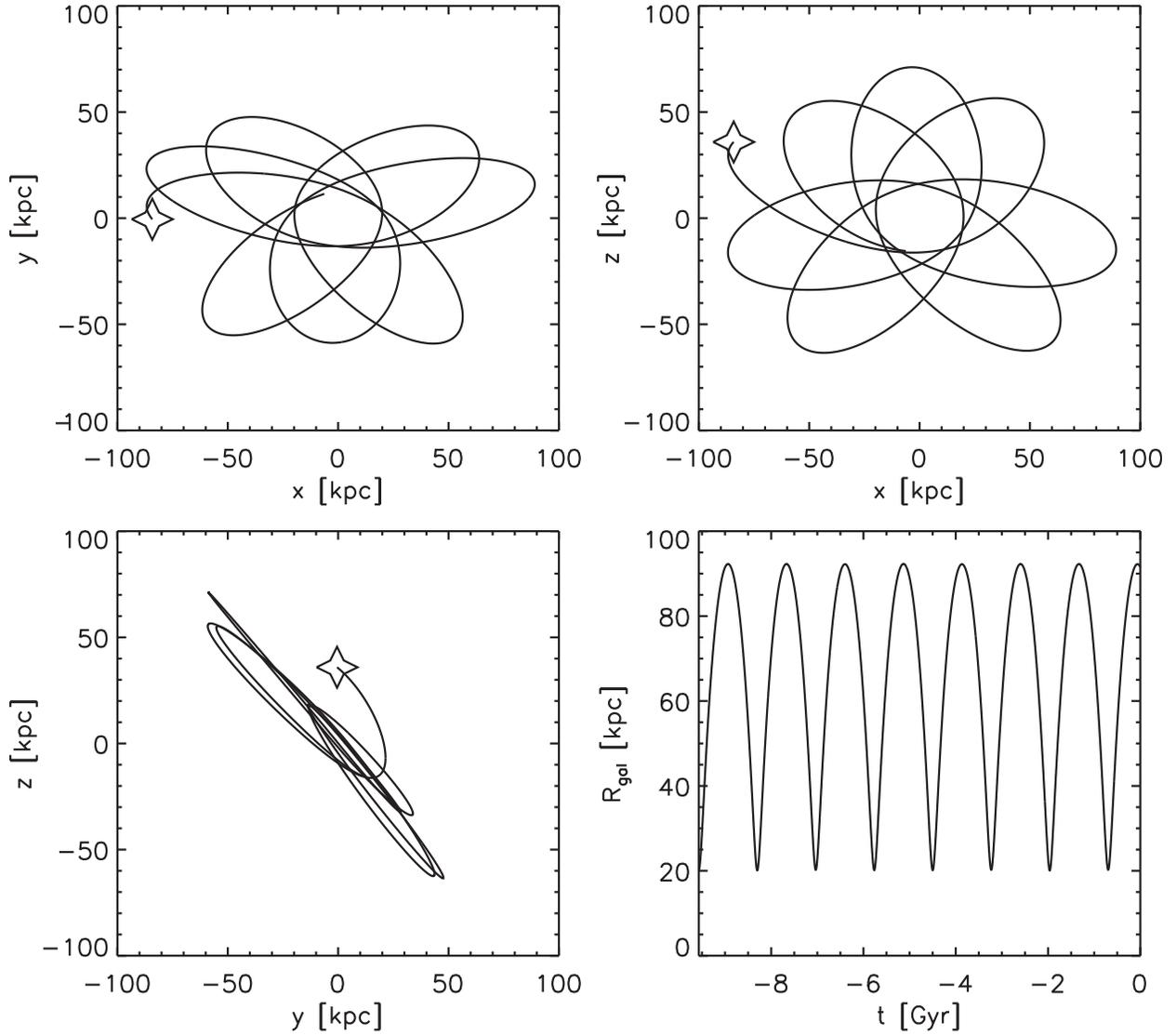}
\caption{The orbit of NGC\,2419 traced back from its current position and projected to the xy-, 
the xz-, and the yz-plane The stars indicate the observed current position of NGC\,2419. 
Lower left: The distance of NGC\,2419 to the Galactic center. The orbital period is about 1.3 Gyr. \label{orbitfig}}
\end{figure}

Figure \ref{orbitfig} shows the orbit of NGC\,2419 calculated back in time using the 
Milky Way potential as given in Sect.~\ref{potential_mw}. The orbital period is about 1.3 Gyr. 
The CC will most likely be formed during the first perigalactic passage of the parent galaxy. 
However, as the calculated objects will not change significantly after about 7 Gyr of evolution 
(see Sect. \ref{timeevolution}), we do not start at $t$ = --12 Gyr (the age of NGC\,2419) in order 
to cut down the computing time. For all numerical simulations we choose the perigalactic passage 
at $t$ = --9.568 Gyr as a starting point. 

\subsection{Initial Configuration of the Cluster Complex and Model Parameters} \label{initial_configuration}

The CC models consist of $N_{\rm 0}^{\rm CC}$ = 20 star clusters and are diced according to a 
Plummer distribution \citep{plum1911, krou08}. The cutoff radius, $R_{\rm cut}^{\rm CC}$, 
of the CC is four times the Plummer radius, $R_{\rm pl}^{\rm CC}$. The initial velocity distribution 
of the CC models is chosen such that the CC is in virial equilibrium. A detailed description of the 
generation of initial coordinates (space and velocity) for Plummer models is given in the appendix 
of \cite{aarseth}.

The individual star clusters building up the CCs in our simulations are Plummer spheres 
with a Plummer radius of $R_{\rm pl}^{\rm SC} = 4$ pc and a cutoff radius of 
$R_{\rm cut}^{\rm SC} = 20$ pc. Each star cluster has a mass of 
$M^{\rm SC} = 0.05 \times M^{\rm CC}$ and consists of $N_{\rm 0}^{\rm SC}$ = 100\,000 particles. 
The velocity distribution of the individual star clusters is chosen to be initially in 
virial equilibrium.

\begin{figure}
\epsscale{0.8}
\plotone{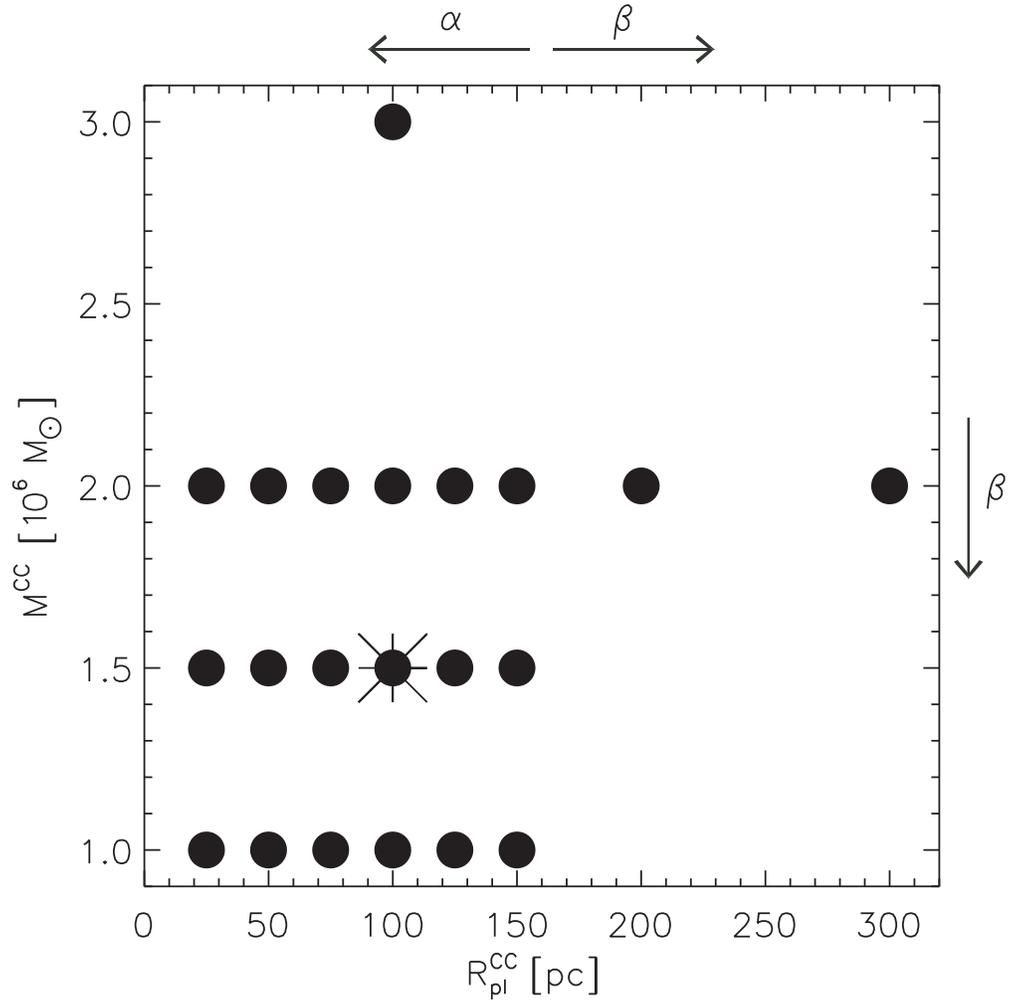}
\caption{The parameter range covered in the $M^{\rm CC}$ vs. $R_{\rm pl}^{\rm CC}$ space. 
The arrows indicate the increase of the parameters $\alpha$ and $\beta$ (see Sect. \ref{massandsize}). 
The circle with the additional asterix marks the values of $M^{\rm CC}$ and $R_{\rm pl}^{\rm CC}$ of 
those models, where the initial distribution of star clusters within the CC was varied 
(see Sect. \ref{variniscdist}). 
\label{figmatrix}}
\end{figure}

\begin{figure}
\epsscale{1.0}
\plotone{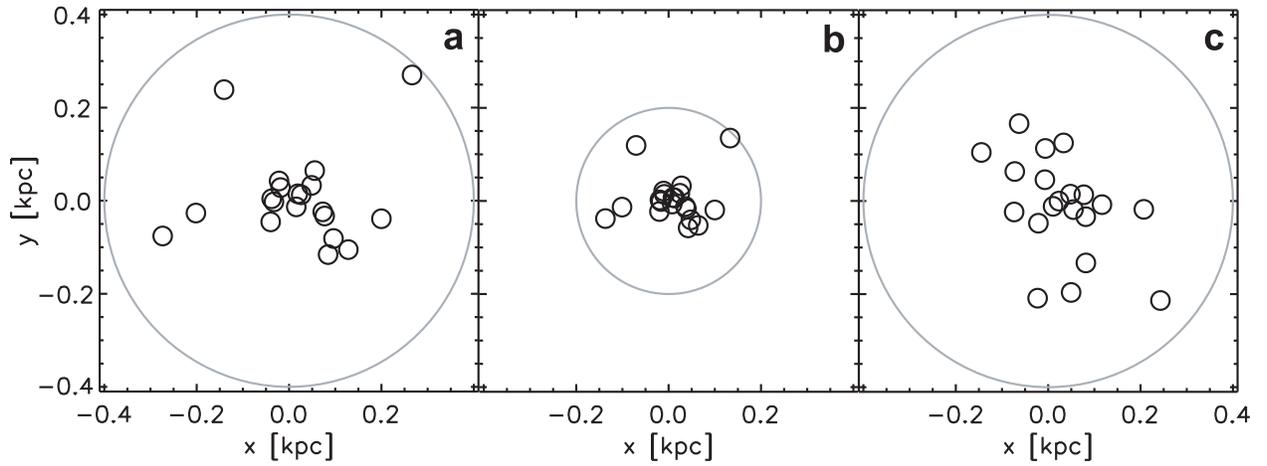}
\caption{Three exemplary initial distributions of star clusters (small circles with radius 
$R_{\rm cut}^{\rm SC}$) in a CC (surrounding circle with radius $R_{\rm cut}^{\rm CC}$) projected 
onto the x-y-plane. Model M\_1\_1.5\_50 (b) is a scaled version of M\_1\_1.5\_100 (a) 
only differing in Plummer radius $R_{\rm pl}^{\rm CC}$. Some star clusters do already overlap 
in the center at the beginning. Model M\_4\_1.5\_100 (c) has a less concentrated distribution
of clusters than M\_1\_1.5\_100. \label{figinimodel}}
\end{figure}

In total, we considered 27 different models (see Tables \ref{tbl-inipar} and \ref{tbl-2}), 
which are denoted M\_$x$\_$y$\_$z$, where $x$ is the number of the initial configuration, 
i.e. the detailed distribution of the individual star clusters in the CC, $y$ is the CC mass, 
$M^{\rm CC}$, in units of 10$^{6}$ M$_{\odot}$, and $z$ is the CC Plummer radius, $R_{\rm pl}^{\rm CC}$, 
in pc. Figure \ref{figmatrix} visualizes the CC parameter range covered in the $M^{\rm CC}$ vs. 
$R_{\rm pl}^{\rm CC}$ space.
Figure \ref{figinimodel} illustrates the different initial distributions. Figure \ref{figinimodel}a and b
are the same initial distribution of star clusters that were scaled according to their 
$R_{\rm pl}^{\rm CC}$, while Figure \ref{figinimodel}c shows a less concentrated distribution of 
star clusters.

\section{RESULTS} \label{results}

We carried out 27 different numerical simulations to get an estimate of the influence 
of varying initial CC conditions. All calculations start at the perigalactic passage at 
$t_{\rm 0}$ = --9.568 Gyr and are calculated up to the current position of NGC\,2419. 

\subsection{Time Evolution of the Merging Process} \label{timeevolution}

\begin{figure}
\epsscale{0.8}
\plotone{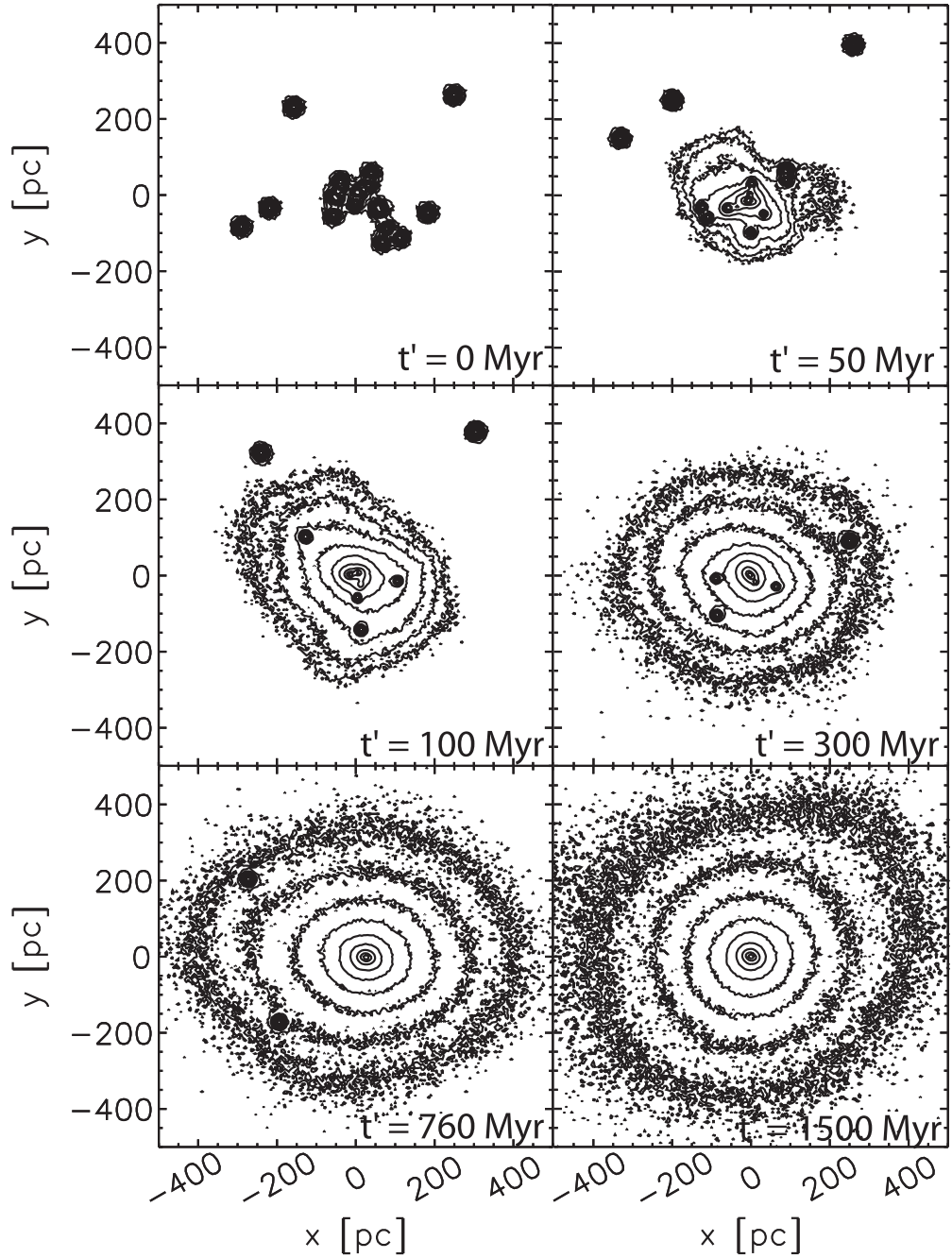}
\caption{Time evolution of the merger object in model M\_1\_1.5\_100. 
Contour plots on the x-y-plane displayed at $t'$ = 0, 50, 100, 300, 760 and 1500 Myr. The lowest contour 
level corresponds to 5 particles per pixel. The pixel size is 5 pc. 
This yields 0.15 M$_{\odot}$ pc$^{-2}$. The contour levels increase further by a factor of 3.
\label{fig_timeevol}}
\end{figure}

The merging process of model M\_1\_1.5\_100 is shown in Figure \ref{fig_timeevol} as contour plots 
on the xy-plane to illustrate the detailed evolution of the merging process. 
The snapshots were taken at $t$' = $t$ -- $t_{\rm 0}$ = 0, 50, 100, 300, 760 and 
1500 Myr. At $t$' = 50 Myr the merger object is already in the process of forming, but the 
majority of star clusters are still individual objects. In the course of time more and more star 
clusters are captured by the merger object. Thus the merger object becomes more extended. After 10 
crossing times ($t$' = 760 Myr) there are still 2 unmerged star clusters in the vicinity of the merger 
object. In the last snapshot at $t$' = 1500 Myr the merging process is completed
and 19 out of 20 star cluster have merged forming a smooth extended object. One star cluster escaped the
merging process. It follows the merger object on its orbit around the Milky Way at a distance of 
about 14 kpc (at $t$' = 9.568 Gyr). 

The timescale of the merging process depends on the initial CC mass, the CC size and the 
distribution of star clusters within the CC. For model M\_1\_1.5\_100, 50 percent of the clusters 
have merged after approximately 100 Myr. The time within which half of the clusters merge
increases with the CC size (from 15 Myr for model M\_1\_1.5\_25 to 200 Myr for model M\_1\_1.5\_150) 
and decreases with CC mass (from 150 Myr for model M\_1\_1.0\_100 to 65 Myr for model M\_1\_3.0\_100).

\begin{figure}
\epsscale{1.0}
\plotone{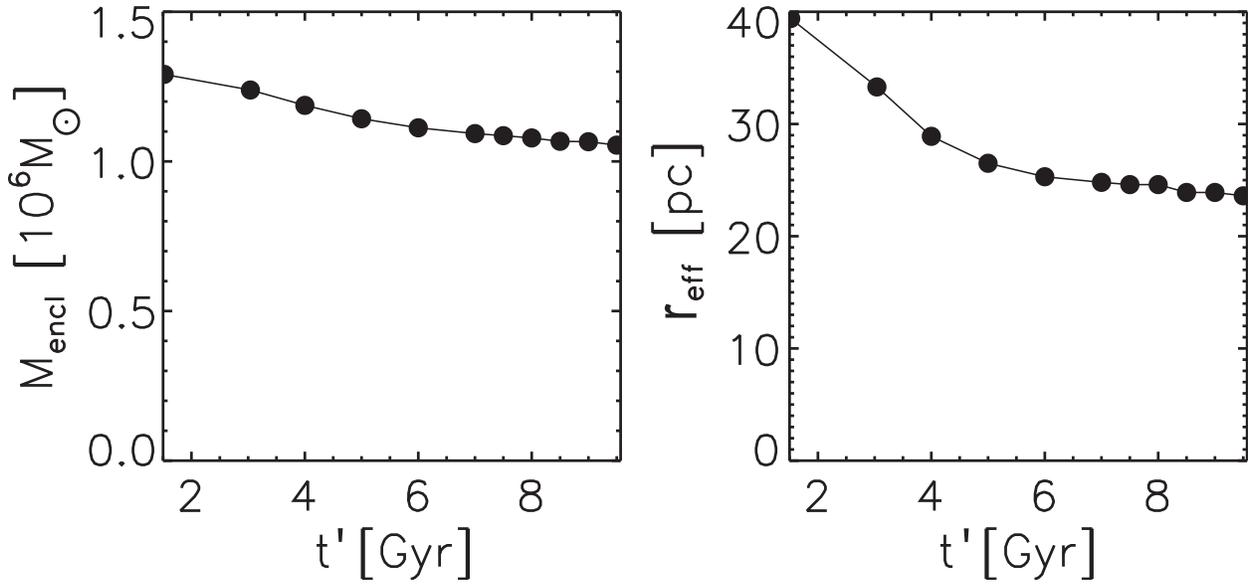}
\caption{
Left: Enclosed mass versus time after 1.5 Gyr of evolution of model M\_1\_1.5\_100. The curve 
becomes fairly flat with increasing time. An almost stable merger object forms suffering only 
slightly from mass loss.
Right: effective radius against time after 1.5 Gyr of evolution of model M\_1\_1.5\_100. The 
effective radius decreases with time up to about 7 Gyr. Thereafter it is almost constant. 
\label{fig_timeevol_M_R}}
\end{figure}

The further time evolution of the merger object of model M\_1\_1.5\_100 is plotted in 
Figure \ref{fig_timeevol_M_R}. The enclosed mass of the merger objects is defined as the 
mass within 800 pc. The half-mass radius is the radius of the sphere within which half of the mass 
is enclosed. However, as observers can only derive a projected half-mass radius, we calculate also a 
value projected on the sky defined as the projected radius within which half of the mass 
is included. The projected half-mass radius is slightly smaller than the three-dimensional half-mass radius 
(Table \ref{tbl-2}) and corresponds to the observed effective radius, $r_{\rm eff}$, if mass follows 
light. The effective radius and mass of the merger object decrease and become fairly 
constant after about 7 Gyr. The structural parameters change only very slightly in the next few Gyr.

\subsection{Variation of initial CC mass and size}\label{massandsize}

We consider models with three CC masses, $M^{\rm CC}$ = 1.0, 1.5, and 2.0 $\times 10^{6}$ 
M$_{\odot}$, and six CC Plummer radii between $R_{\rm pl}^{\rm CC}$ = 25 and 150 pc in steps of 25 pc 
to analyze the dependence of the structural parameters of the merger objects on the initial CC mass and size. 
In addition, two models with $R_{\rm pl}^{\rm CC}$ = 200 and 300 pc are calculated for 
$M^{\rm CC}$ = 2.0 $\times~10^{6}$. All models have the same relative initial distribution of star clusters, 
which is scaled according to the respective Plummer radii (see Figure \ref{figinimodel}a and b). The 
velocities of the individual star clusters are also scaled with CC mass and size to keep the CCs 
initially in virial equilibrium.

For all models the merging process leads to a stable object. The number of merged star clusters
is between 10 and 20. The properties of the merger objects at the current position of NGC\,2419 are 
displayed in Table \ref{tbl-2}.

Figure \ref{figmassreff}a shows the enclosed mass, $M_{\rm encl}$, of the merger objects
as a function of the initial CC Plummer radius, $R_{\rm pl}^{\rm CC}$. The fraction of the initial CC mass, 
which is bound to the merger object, decreases almost linear with increasing CC size from about 92\% 
at $R_{\rm pl}^{\rm CC}$ = 25 pc for all three masses to values of 46, 53, and 59\%  at 
$R_{\rm pl}^{\rm CC}$ = 150 pc for CC masses of $M^{\rm CC}$ = 1.0, 1.5, and 2.0 $\times~10^{6}$, respectively. 
Mass-loss occurs either by the escape of individual stars from the diffuse stellar component, 
which builds up during the merging process (see Sect. \ref{timeevolution}), or by entire star 
clusters escaping the merging process. For all three masses, all 20 star clusters merge
for compact models up to $R_{\rm pl}^{\rm CC} $ = 75 pc, i.e. mass-loss is only from the diffuse 
component for these models. For the more extended models with $R_{\rm pl}^{\rm CC} \ge$  100 pc 
star clusters escape from the merging process. For extended models with $R_{\rm pl}^{\rm CC}$ = 150 pc 
six, five, and two star clusters escape for CC masses of $M^{\rm CC}$ = 1.0, 1.5, and 2.0 $\times~10^{6}$, 
respectively.

Figure \ref{figmassreff}b shows the effective radii, $r_{\rm eff}$, of the merger objects
as a function of the initial CC Plummer radius, $R_{\rm pl}^{\rm CC}$. The effective radii of the 
merger objects increase with increasing CC size up to $R_{\rm pl}^{\rm CC}$ = 75 pc for 
$M^{\rm CC}$ = 1.0~$\times$~10$^{6}$ and up to $R_{\rm pl}^{\rm CC}$ = 100 pc for the more massive 
models. For larger values of $R_{\rm pl}^{\rm CC}$ the effective radii decrease rapidly. 
Despite the large range of CC sizes ($R_{\rm pl}^{\rm CC}$ = 25 to 300 pc), 
the effective radii of the merger objects are constrained to a relatively small interval, e.g. models 
with $M^{\rm CC}$ = 2.0~$\times$~10$^{6}$ are constrained to values between 13.2 and 25.6 pc.

Figure \ref{turnover} combines Figures \ref{figmassreff}a and b showing the effective 
radius of the merger objects as a function of their enclosed mass. Only the six radii 
$R_{\rm pl}^{\rm CC}$ = 25, 50, 75, 100, 125, and 150 pc, where data for all three masses are available, 
were used to allow for an overview on the trends (models with the same initial CC Plummer radius are 
connected by dotted lines, models with the same initial mass with solid lines).
For all CC Plummer radii the effective radii increase with the initial CC mass. 
While the effective radii vary only slightly with mass for models with $R_{\rm pl}^{\rm CC}$ = 25, 50 
and 75 pc, the slope is considerably steepening for larger CC sizes.

The formation process of the merger objects depends on the compactness of the initial CC. 
A measure of how densely a CC is filled with star clusters for an equal number $N_{\rm 0}^{\rm CC}$ 
of star clusters is given by the parameter $\alpha$ \citep{fell02a},
\begin{eqnarray} \alpha = \frac{R_{pl}^{SC}}{R_{pl}^{CC}}, \end{eqnarray}                                                                                                        
where $R_{\rm pl}^{\rm SC}$ and $R_{\rm pl}^{\rm CC}$ are the Plummer radius of a single star cluster 
and the Plummer radius of the CC, respectively. In our models, the individual star clusters are not 
scaled with $R_{\rm pl}^{\rm CC}$. This leads to decreasing values of $\alpha$ from 0.16 for 
$R_{\rm pl}^{\rm CC}$ = 25 pc to $\alpha$ = 0.013 for $R_{\rm pl}^{\rm CC}$ = 300 pc. In general, large 
values of $\alpha$ accelerate the merging process because the star clusters already partly overlap 
in the center of the CC, whereas small values hamper the merging process.

Also the tidal field counteracts the merging process. An estimate of the influence of the tidal 
field on the CC is given by the parameter 
\begin{eqnarray} \beta =\frac{R_{cut}^{CC}}{r_{t}^{CC}} \end{eqnarray} \citep{fell02a}, 
which is the ratio of the cutoff radius $R_{\rm cut}^{\rm CC}$ of the CC and 
its tidal radius $r_{\rm t}^{\rm CC}$. The tidal radius, $r_{\rm t}^{\rm CC}$, is defined as 
the radius where the attractive force of the CC on an object equals the pulling force 
of the Milky Way. Using the Milky Way potential as given in Section \ref{potential_mw} and the 
CC potential, which is given by the Plummer potential
\begin{eqnarray} 
\Phi^{\rm CC}_{\rm pl}(r) = -G \frac{M^{\rm CC}}{R_{\rm pl}^{\rm CC}\sqrt{(1+\frac{r}{R_{\rm pl}^{\rm CC}})^2}},
\end{eqnarray}
the tidal radius at the peri-galactic distance of 20 kpc is $r_{\rm t}^{\rm CC}$ = 296, 291, 283, 274, and 262 pc 
for $M^{\rm CC}$ = 1.5 $\times$ 10$^{6}$ and $R_{\rm pl}^{\rm CC} =$ 50, 75, 100, 125 and 
150 pc, respectively. The corresponding values of $\beta$ are 0.68, 1.03, 1.41, 1.83 and 2.29.
The values of $\beta$ are slightly higher for the models with $M^{\rm CC}$ = 1.0~$\times$ 10$^{6}$ 
and slightly lower for models with $M^{\rm CC}$ = 2.0~$\times$ 10$^{6}$ (Table \ref{tbl-beta}).
The fraction of the enclosed mass of a merger object and the initial CC mass, $\frac{M_{\rm encl}}{M^{\rm CC}}$, 
is plotted as a function of $\beta$ in Figure \ref{massandbeta}. The merger objects of the models for 
the three initial CC masses show the same linear correlation with $\beta$. 

The models with $R_{\rm pl}^{\rm CC}$ = 75 pc start already at the peri-galactic distance
with a $\beta$ of about one. Consequently, all 20 clusters merge. For the more extended models
a number of star cluster are initially located outside the tidal radius. As we use a highly eccentric 
orbit, the CCs move rapidly towards larger Galactic distances, e.g. after 100 Myr they are at a distance 
of about 35 kpc. Due to the lower gravitational field of the Milky Way at larger distances, the tidal 
radii increase leading to lower values of $\beta$. Figure \ref{betaplot} shows the variation of $\beta$
with time for models with $M^{\rm CC}$ = 1.5 $\times$ 10$^{6}$. For models M\_1\_1.5\_100, M\_1\_1.5\_125, 
and M\_1\_1.5\_150 the period with $\beta >$ 1 is getting longer, resulting in lower numbers of merged 
clusters for larger CCs. The turnover in $r_{\rm eff}$, as shown in Figures \ref{figmassreff}b 
and \ref{turnover}, occurs at those $R_{\rm pl}^{\rm CC}$, where $\beta$ is sufficiently large to 
allow entire star clusters to escape the merging process. 

\subsection{Variation of the initial distribution of star clusters}\label{variniscdist}
We use the CC parameters of model M\_1\_1.5\_100 as a basis to analyze the influence of the detailed
distribution of star clusters in the CC. We calculate the evolution of five additional models, of which two
(M\_2\_1.5\_100 and M\_3\_1.5\_100) have a similar concentration of clusters in their center as 
M\_1\_1.5\_100, whereas the other three models (M\_4\_1.5\_100, M\_5\_1.5\_100 and M\_6\_1.5\_100) 
show a less concentrated distribution of star clusters (see Figure \ref{figinimodel}c). All six models
result from exactly the same Plummer model but with different random number seeds.

The average and the standard deviation of the effective radii and enclosed masses of the
merger objects of all six models are $r_{\rm eff}$ = 23.8$\pm$3.2 pc and 
$M_{\rm encl}$ = 0.95$\pm$0.17 $\times$ 10$^{6}$ M$_{\odot}$, respectively. The standard deviations, 
which correspond to relative deviations of 13\% ($r_{\rm eff}$) and 18\% ($M_{\rm encl}$), provide an
order of magnitude estimate of the influence of the distribution of star clusters on the structural
parameters of the merger objects.

The merger objects resulting from compact initial configurations 1 to 3 have on average a 
significantly higher mass 
($M_{\rm encl}$ = 1.10$\pm$0.06 $\times$ 10$^{6}$ M$_{\odot}$) than the less concentrated configurations 4 to 6
($M_{\rm encl}$ = 0.81$\pm$0.06 $\times$ 10$^{6}$ M$_{\odot}$). In contrast, no clear difference in
the effective radii of the merger objects can be seen between concentrated and extended initial distributions.

For comparison, an additional model M\_7\_1.5\_200 has been calculated. It has a CC Plummer 
radius of 200 pc and a relatively small cutoff radius of 400 pc. A scaled version of model 
M\_1\_1.5\_100 with $R_{\rm pl}^{\rm CC}$ = 200 pc and a CC cutoff radius of 800 pc, which is four times the 
CC Plummer radius, would have 6 star clusters beyond 400 pc. Model M\_7\_1.5\_200 has the same cutoff 
radius and therefore the same values of $\beta$ as model M\_1\_1.5\_100. Due to the broader initial 
distribution of star cluster in M\_7\_1.5\_200 the enclosed mass and the effective radius are slightly 
lower than in model M\_1\_1.5\_100. 

The results demonstrate that next to the CC mass and the CC Plummer radius also the 
cutoff radius and the exact distribution of star clusters in a CC are key parameters for 
extended models, considerably influencing the structural parameters of the merger objects.

\subsection{Comparison with Observations}\label{comp_to_obs}

The enclosed masses were converted into absolute $V$-magnitudes to allow 
for direct comparison with the observed data, using the formula
\begin{eqnarray}
M_{\rm V} = M_{\rm V,solar} - 2.5 \times \log(M_{\rm encl}\frac{L_{\rm V}}{M}), 
\end{eqnarray}
where, $M_{\rm V,solar}$ = 4.83 mag is the absolute solar $V$ magnitude, $M_{\rm encl}$ the 
enclosed mass of the merger object and $\frac{M}{L_V}$ the mass-to-light ratio. 
We use a mass-to-light ratio of 2.05$\pm$0.50, as determined by \cite{baumgardt} for NGC\,2419. 
Apparent V-magnitudes can by calculated by adding the distance modulus of NGC\,2419 (m-M)$_0$ = 19.60 
\citep{ripepi} and a V-band extinction of $A_{\rm V}$ = 0.25 using a reddening of E(B-V) = 0.08 
\citep{ripepi}. 

Figure \ref{contourplot} shows an exemplary contour plot of the merger object of model
M\_1\_1.5\_100 projected onto the sky using Galactic coordinates. The lowest contour 
line corresponds to 32 mag arcsec$^{-2}$.
At this low surface brightness, NGC\,2419 is detectable up to radii of about 17\arcmin\ corresponding 
to 415 pc. Observed photometry, however, covers only a region within the inner 60\arcsec\ of 
NGC\,2419, while surface densities need to be derived from star counts in the outer parts. 
\cite{ripepi} found evidence for stars associated with NGC\,2419 up to radii of 15\arcmin.

Figure \ref{obssims} shows the effective radius, $r_{\rm eff}$, versus the absolute magnitude,
$M_{\rm V}$, of the observations and the models as given in Table \ref{tbl-1} and \ref{tbl-2}, 
respectively.
Due to the scatter of the observed parameters of NGC\,2419 and the turnover in the effective radii 
of the models as described in Section \ref{massandsize}, a large number of models are compatible with 
the observed parameters of NGC\,2419. Only extended models with initial CC masses of 
$M^{\rm CC}$ = 1.0~$\times$ 10$^{6}$
and compact models with $M^{\rm CC}$ = 2.0~$\times$ 10$^{6}$ are clearly incompatible with observations.

A more detailed comparison between models and observations can be achieved using surface brightness 
profiles. \cite{bellazzini} compiled a surface brightness profile out to projected radii of 
477\farcs5. However, as he used radii between 720\arcsec\ and 900\arcsec, 
which are well inside NGC\,2419, as a reference field to estimate the background level, 
Bellazini's surface brightness estimates are expected to be slightly 
too faint in the outer parts NGC\,2419.

The radial surface brightness profiles of four exemplary models projected 
onto the sky and converted to units of mag arcsec$^{-2}$ are shown in Figure \ref{surfbright}. 
The merger objects show a King-like profile out to radii of about 1000\arcsec. 
The observed profile from \cite{bellazzini} is added to Figure \ref{surfbright} to allow for a 
direct comparison. The surface brightness profile of model M\_1\_1.0\_50 agrees very well
with the observed profile at all radii. The other three models in Figure \ref{surfbright} illustrate
how the profiles change when one of the parameter mass, size and initial distribution of star clusters
is modified. Model M\_1\_1.5\_50 has a similar shape as M\_1\_1.0\_50, but is too bright at all radii 
due to the larger mass. Model M\_1\_1.5\_100, which is a more extended version of model M\_1\_1.5\_50, 
agrees well with the observations between 10\arcsec and 100\arcsec, but it is considerably brighter in 
the center and the outer parts. Model M\_4\_1.5\_100, which has a broader initial distribution of star 
clusters than model M\_1\_1.5\_100, shows a smaller deviation from the observed surface brightness profile 
than model M\_1\_1.5\_100.

\cite{baumgardt} observed the radial velocities of 40 stars within a projected radius of 100 pc of 
NGC\,2419 and derived a velocity dispersion of $\sigma$ = 4.14$\pm$0.48 km s$^{-1}$. The line-of-sight 
velocity dispersions within a projected radius of 100 pc of the models are listed in 
Table \ref{tbl-2}. A number of models with masses $M^{\rm CC}$ = 1.0, 1.5, and 2.0 $\times$ 10$^{6}$ M$_{\odot}$
have velocity dispersions that are within the one sigma error of the observed velocity dispersion.
Model M\_1\_1.0\_50, which has an effective radius and enclosed mass very close to the newest
observed values from \cite{baumgardt} and a surface brightness profile that is a good approximation 
of the observed profile, has a velocity dispersion of $\sigma$ = 4.13 km s$^{-1}$, i.e. almost exactly
the observed one.

Considering masses, effective radii, surface brightness profiles, and velocity dispersions, 
model M\_1\_1.0\_50 provides the best representation of NGC\,2419. However, a number of 
models reproduce the observed structural parameters of NGC\,2419 quite well within the
observational uncertainties, demonstrating that an object like NGC\,2419 
can be formed from merged CCs without the need of fine-tuning of the input parameters.

\section{DISCUSSION} \label{discussion}

\subsection{The Merged Cluster Complex Scenario}

The proposed formation scenario for NGC\,2419 starts with newly born complexes of star clusters
in the Galactic halo with orbital parameters allowing for an highly eccentric orbit. We model the 
dynamical evolution of various CCs leading to merger objects. We do not, however, consider the 
process which formed the CCs in the first place as this would increase the complexity of the simulations 
and add more degrees of freedom making the interpretation of the results difficult. 

For all 27 models the majority of star clusters merge into a stable object. The turnover 
in the $r_{\rm eff}$ vs. $M_{\rm encl}$ space (Figure \ref{figmassreff}, \ref{turnover} and \ref{obssims}) 
leads to degenerate states, because a relatively compact CC can produce a comparable merger object as a 
more massive CC having a significantly larger CC size. In consequence, a range of initial conditions can 
form a merger object comparable to NGC\,2419 preventing us to pinpoint the parameters of the original CC, 
which formed NGC\,2419. On the other hand, the larger the range of initial conditions that end up 
in a NGC\,2419-like object, the larger is the probability of creating a massive EC like NGC\,2419.

As the individual star clusters of the CC formed at approximately the same
time from molecular clouds of a galaxy, the observed absence of multiple stellar populations 
\citep{ripepi} is fully consistent with our model. A small scatter in metallicity would 
also be explained, as the individual pre-cluster cloud cores of the complex could have had slightly 
different metallicities.
In addition, some stars from the parent galaxy might have been captured by NGC\,2419 during 
its formation as demonstrated for massive star clusters like $\omega$ Cen by \cite{fellhauer06}.

Observations demonstrate that massive complexes of star clusters nowadays predominantly
form during gravitational encounters between late-type galaxies. Young massive CCs 
with masses above 10$^6\,M_\odot$ and sizes of a few hundred pc, comparable to those used as initial
conditions for the numerical simulations presented in Sect. \ref{results}, have been observed in the
Antennae galaxies \citep{bastian06} and in NGC\,922 \citep{pellerin}.

Since galaxy-galaxy mergers are anticipated to have been much more common during 
early hierarchical structure formation it is expected that star formation in cluster complexes has been 
a significant star formation mode during early cosmological epochs. 

\subsection{Potential Association with Stellar Streams in the Milky Way halo}

As stellar structures in the outer halo are long lived features, remnants of a parent galaxy or 
stellar tidal streams may still be observable in the Milky Way halo. 

\cite{newberg03} found an over-density of A-type stars at a distance of 83 to 85 kpc, which has a 
width of at least 10$^\circ$ and which was traced for more than 20$^\circ$ on the sky. NGC\,2419 
is located within this debris (on the sky and at the same distance). \cite{newberg03} argued that 
NGC\,2419 and the stellar over-density might be associated with the Sagittarius dwarf galaxy, as 
NGC\,2419 lies only 13 kpc from its orbital plane. However, the stellar streams from Sgr dwarf 
have mean metallicities between [Fe/H] = --0.4 near the core and --1.1 within the arms \citep{chou07}, 
while NGC\,2419 has a very low metallicity of [Fe/H] = --2.12 \citep{harris}.

The Virgo Stellar Stream (VSS) has a metallicity of [Fe/H] = --1.86 with a large scatter of 
0.40 dex \citep{duffau}, which makes it comparable to the metallicity of NGC\,2419. \cite{casetti} 
determined the proper motion of the VSS and calculated an orbit with a pericentric distance of 11 kpc 
and an apocentric distance of 89 kpc. They concluded that the current position of NGC\,2419 is 
compatible with this orbit.

\cite{newberg09} found a new stellar stream at a distance of 35 kpc in the constellation Cetus having a 
nearly polar orbit. The so-called Cetus Polar Stream (CPS) has a very low metallicity of [Fe/H] = --2.1.
While the metallicity of CPS is in excellent agreement with NGC\,2419, the polar orbit along Galactic 
longitude $l = 143^\circ$ appears to be in conflict with the current location of NGC\,2419, which is 
at $l = 180^\circ$.

The Orphan Stream, independently discovered by \cite{grillmair} and \cite{belekurov}, has been analyzed 
in detail by \cite{newberg10}. They found a very low metallicity of [Fe/H] = --2.1 and a likely orbit 
between a pericentric distance of 16.4 kpc and an apocentric distance of about 90 kpc. These parameters 
are very similar to those of NGC\,2419. 

Any potential association with the Sgr stream, VSS, CPS, or Orphan stream can only be verified 
after the proper motion of NGC\,2419 has been measured. 

\subsection{NGC\,2419-like Objects in other Galaxies}

NGC\,2419 is located in the outer Galactic halo. Also the lower mass ECs of the Milky
Way are halo objects: 9 out of 13 ECs have galacto-centric distances greater than 20 kpc 
\citep{harris}. A similar trend has been shown for the other two disk galaxies in the Local 
Group: 12 out of 13 ECs associated with M31 and both ECs found in M33 have projected radii well 
outside the optical disks of these galaxies \citep{huxor08,stonkute,huxor09}. No 
EC as massive as NGC\,2419 has been found in these galaxies.

Due to the limited field of view of the Hubble Space Telescope, most extragalactic 
studies on GCs cover only (a part of) the optical disk of the respective galaxies. An outer 
halo object like NGC\,2419 would be found only by chance if it is projected onto the main 
body of a galaxy. Another reason for incompleteness is the difficulty of distinguishing ECs 
from background galaxies. Therefore, a number of surveys applied a size limit to reduce the 
contamination of background galaxies, e.g. the large GC surveys of the Virgo Cluster \citep{jordan05} 
and the Fornax Cluster \citep{masters10}, covering dozens of galaxies. Their size limit of 
$r_{\rm eff} <$ 10 pc excludes all ECs from their GC catalogues.

While NGC\,2419-like objects with masses of $M_{\rm EC} \approx 10^6$ M$_\odot$ were 
not in the focus of extra-galactic surveys, much effort has been made to detect and to analyze 
stellar objects of $M \approx 10^7$ M$_\odot$ since the discovery of ultra-compact dwarf 
galaxies (UCDs) in the Fornax Cluster by \cite{hilker99} and \cite{drinkwater00}. A number of 
massive stellar objects with effective radii of $r_{\rm eff} \approx$ 20 pc have been found
in the Fornax and the Virgo cluster \citep{mieske08,hasegan,evstigneeva07}.

\cite{richtler} present a NGC\,2419-like object in the halo of the elliptical galaxy 
NGC\,1399 which is the central galaxy of the Fornax Cluster. This object, labeled 90:12 
in \cite{richtler}, is located at a projected distance of about 40 kpc and has a very blue color 
indicating a very low metallicity. It has an absolute V-band magnitude of $M_{\rm V}$ = $- 10.04$ mag,
which corresponds to a mass of $M_{\rm 90:12}$ = 1.8 $\times$ 10$^{6}$ M$_{\odot}$, assuming the 
same mass-to-light ratio as NGC\,2419. The effective radius of $r_{\rm eff} =$ 27 pc is larger
than that of NGC\,2419. In Figure \ref{obssims}, this EC is located between the values of models 
M\_1\_2.0\_100 and M\_1\_3.0\_100. Therefore, the parameters of EC 90:12 are consistent with being
a more massive version of NGC\,2419.

Considerably larger and more complete datasets of stellar halo objects like ECs and UCDs 
are necessary to draw statistically significant conclusions on the question whether NGC\,2419 
is a low-mass UCD, or a high-mass EC, and whether they are all remnants of merged CCs. 
A parametric study covering the entire mass range from ECs to UCDs 
($M^{\rm CC} = 10^{5.5}$ to $10^{8}$ M$_\odot$) will be presented in a subsequent paper 
\citep{bruens11}.

\section{SUMMARY} \label{summary}

The Galactic globular cluster NGC\,2419 has unique parameters. It is one of the most luminous,
one of the most distant, and as well one of the most extended GCs of the Milky Way.
Apart from these unusual parameters, NGC\,2419 appears to be a normal Galactic GC having a low 
metallicity and a single stellar population.

We propose a new formation scenario for NGC\,2419, being a remnant of a merged star cluster 
complex, which was formed during an interaction between a gas-rich 
galaxy and the Milky Way.  To test this hypothesis, we performed 
particle-mesh-code computations of 27 different CC models. We vary the CC mass, the CC size, 
and the initial distribution of star clusters in the CC to analyze the influence of these 
parameters on the resulting objects. These CCs are comparable to those observed in the 
Antennae galaxies \citep{bastian06} and in NGC\,922 \citep{pellerin}.

For all 27 models, the vast majority of star clusters merged into a stable object. 
The derived parameters mass, absolute V-band magnitude, effective radius, velocity dispersion 
and the surface brightness profile are, for a number of models, in good agreement with those 
observed for NGC\,2419.

\begin{figure}
\epsscale{0.7}
\plotone{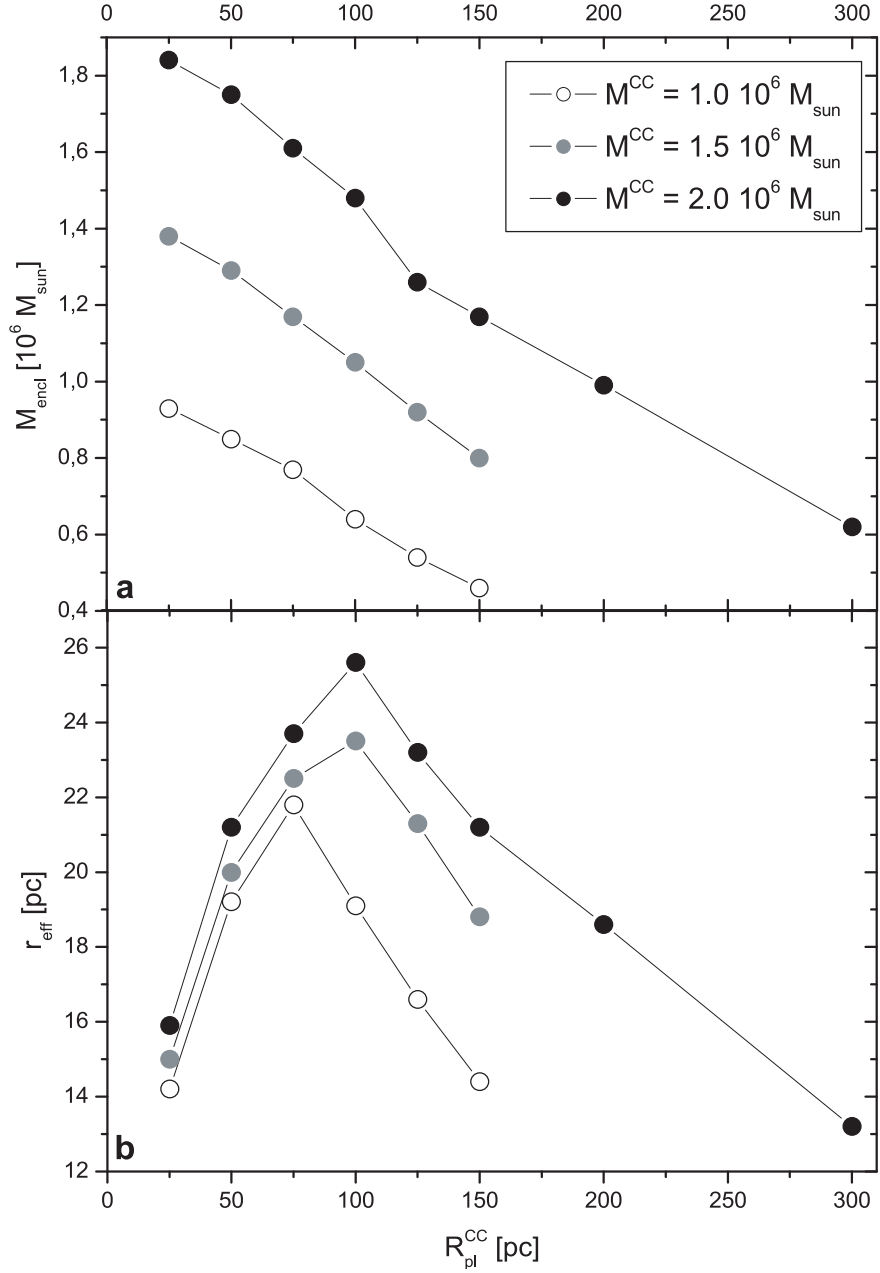}
\caption{\textbf{a}: Enclosed mass, $M_{\rm encl}$, of the merger objects as a function of the
initial CC Plummer radius, $R_{\rm pl}^{\rm CC}$. \textbf{b}: Effective radius, $r_{\rm eff}$,
of the merger objects as a function of the initial CC Plummer radius, $R_{\rm pl}^{\rm CC}$. 
\label{figmassreff}}
\end{figure}

\begin{figure}
\epsscale{1.0}
\plotone{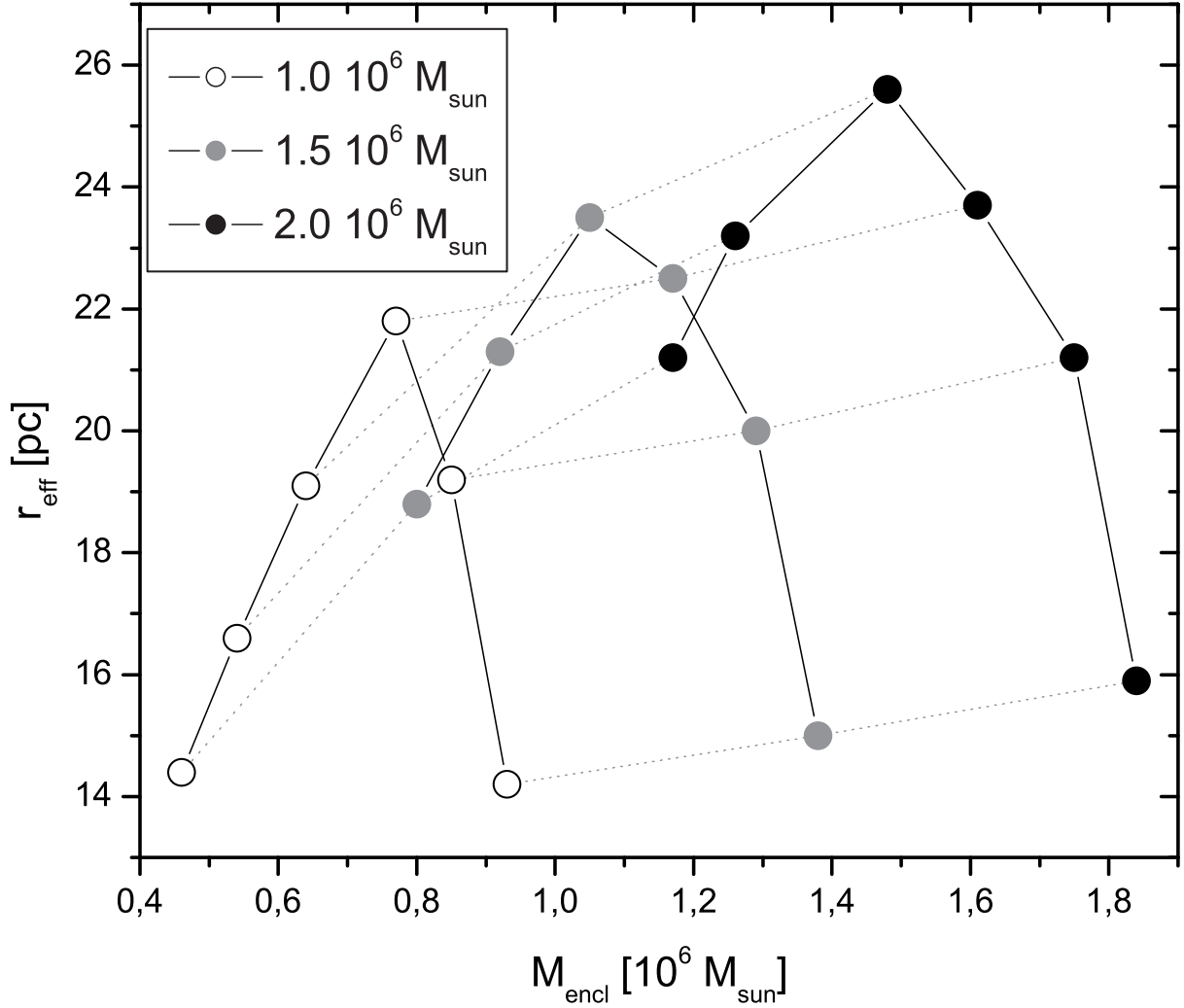}
\caption{Effective radius $r_{\rm eff}$ as a function of the enclosed mass of the merger objects 
for the CC models with $R_{\rm pl}^{\rm CC}$ = 25, 50, 75, 100, 125, and 150 pc (symbols from right to left)
and $M^{\rm CC}$ = 1.0, 1.5, and 2.0 $\times$ 10$^{6}$ M$_{\odot}$. Models with the same initial CC mass 
are connected by solid lines. Models with the same $R_{\rm pl}^{\rm CC}$ are connected by dotted 
lines to illustrate the dependence on the initial CC mass.
\label{turnover}}
\end{figure}

\begin{figure}
\epsscale{1.0}
\plotone{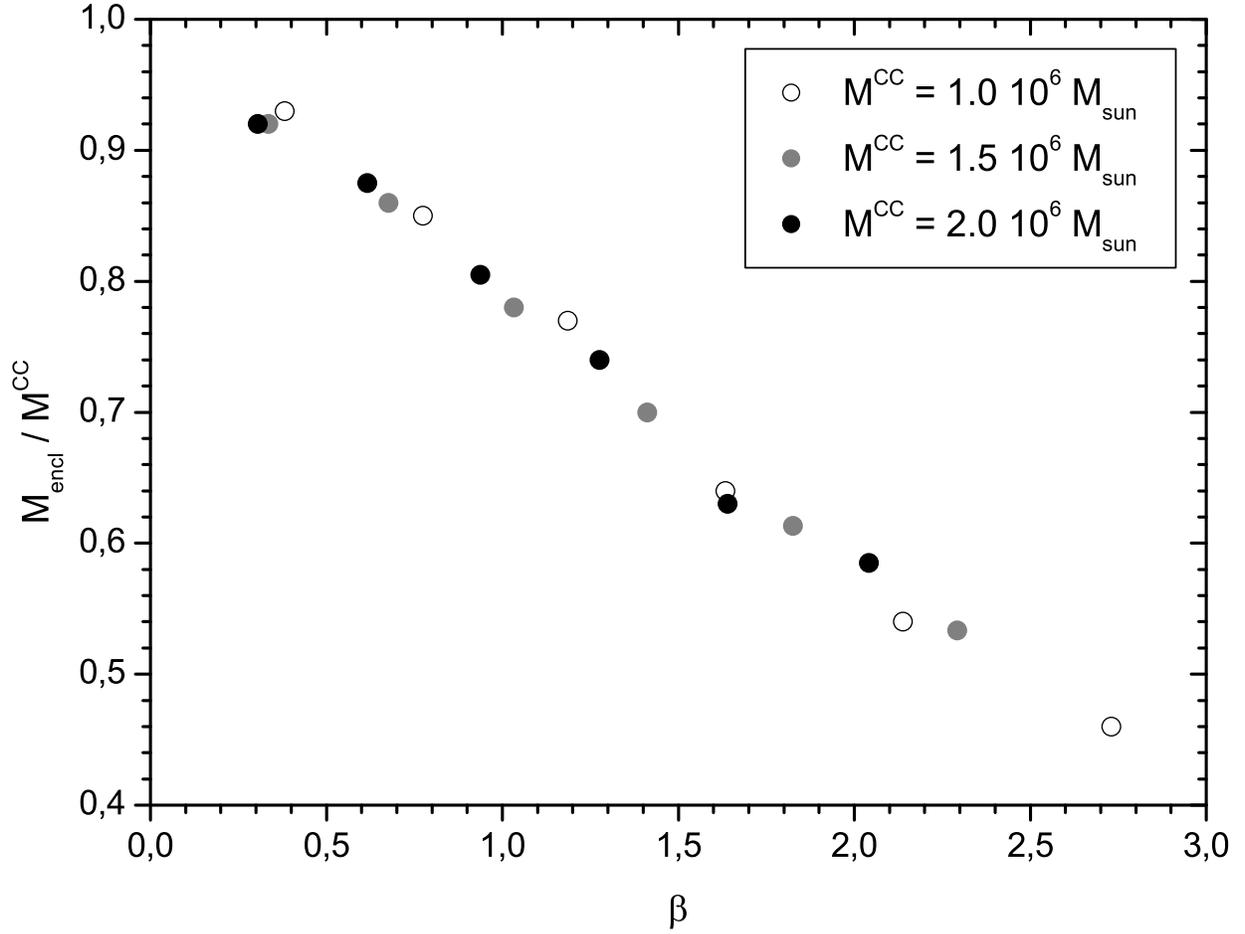}
\caption{The ratio of the merged mass and the initial CC mass, $M_{\rm encl}$/$M^{\rm CC}$, 
as a function of the parameter $\beta$ for the models with $R_{\rm pl}^{\rm CC}$ = 25, 50, 75, 
100, 125, and 150 pc. 
\label{massandbeta}}
\end{figure}

\begin{figure}
\epsscale{1.0}
\plotone{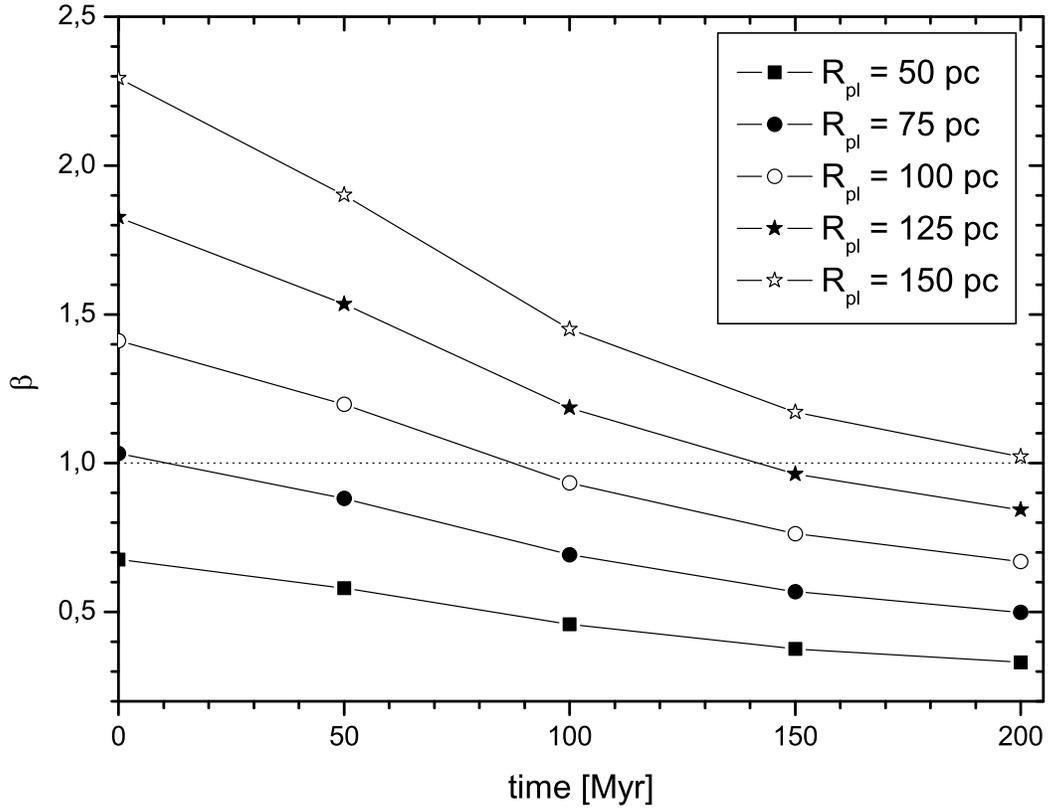}
\caption{Parameter $\beta$ as a function of time for models with 
$M^{\rm CC}$ = 1.5 $\times$ 10$^{6}$ M$_{\odot}$ and $R_{\rm pl}^{\rm CC}$ = 50, 75, 100,
125, and 150 pc for the first 200 Myr (the orbital period is about 1.3 Gyr). \label{betaplot}}
\end{figure}

\begin{figure}
\epsscale{1.0}
\plotone{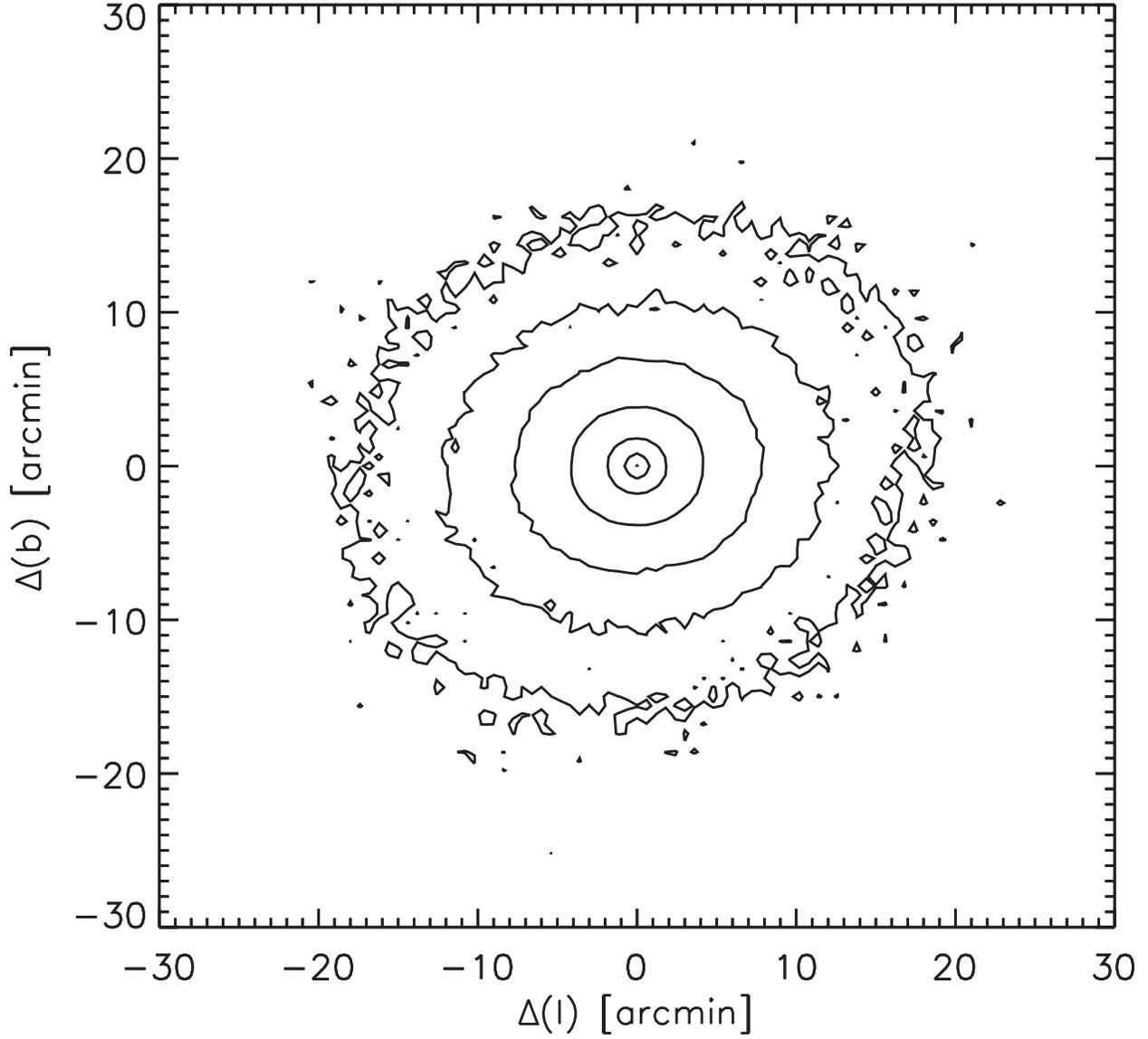}
\caption{Contour plot projected onto the sky in Galactic coordinates for model M\_1\_1.5\_100 at the 
current observed position of NGC\,2419. The pixel size is 0\farcm6. 
The contour levels go from 32 to 20 mag arcsec$^{-2}$ in steps of two mag arcsec$^{-2}$.\label{contourplot}}
\end{figure}

\begin{figure}
\epsscale{1.0}
\plotone{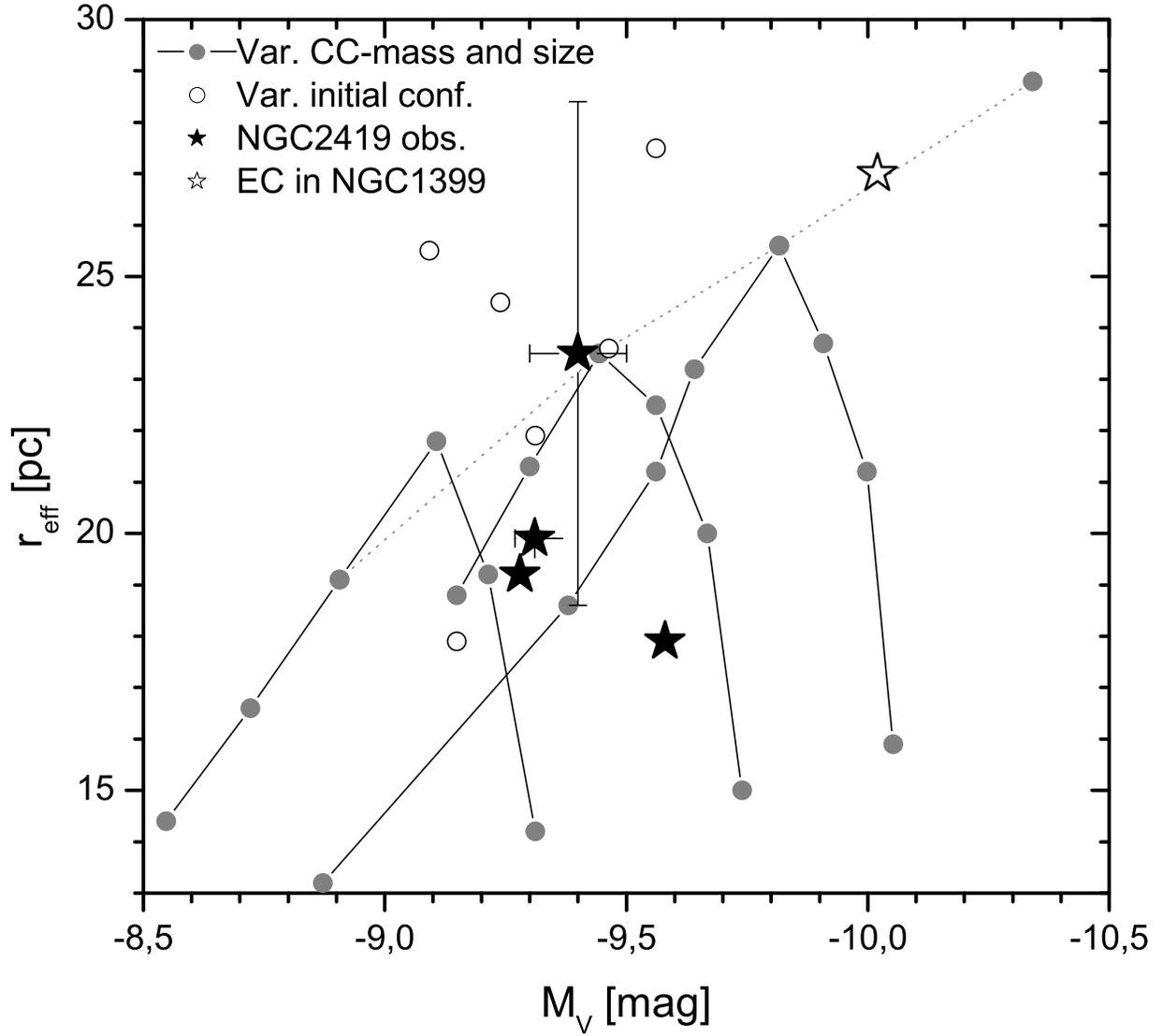}
\caption{Effective radius, $r_{\rm eff}$, as a function of absolute magnitude, M$_{\rm V}$, 
for all computations (see Table~\ref{tbl-2}, the dotted line connects models with $R_{\rm pl}^{\rm CC}$ = 100 pc) 
and observed values of NGC\,2419 (black stars). 
The error bars for the observed values are from the respective papers (see Table~\ref{tbl-1}). 
In addition, EC 90:12 from NGC\,1399 \citep{richtler} is included (white star). \label{obssims}}
\end{figure}

\begin{figure}
\epsscale{0.7}
\plotone{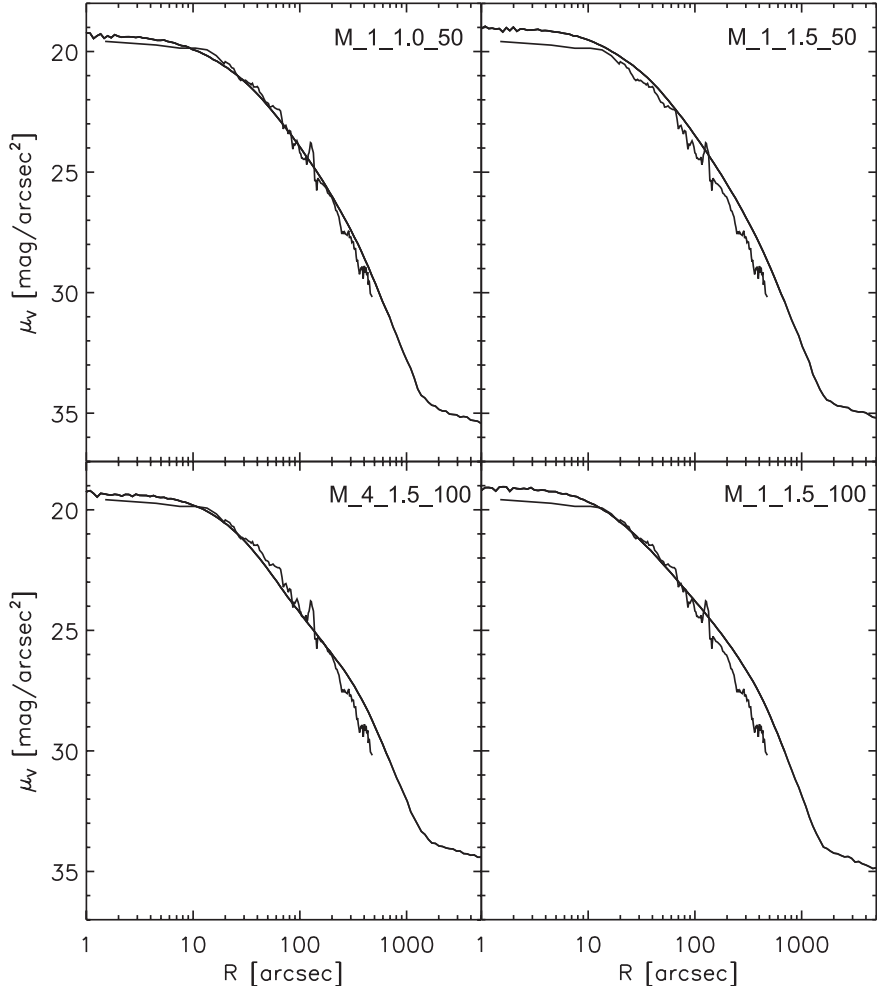}
\caption{Surface brightness profiles of four exemplary models at the current position of NGC\,2419 using a
mass-to-light ratio of 2.05. A V-band extinction of $A_{\rm V}$ = 0.25 was applied to the models to allow 
for a direct comparison with the observed surface brightness profile of \cite{bellazzini}. 
\label{surfbright}}
\end{figure}

The effective radii of the merger objects increase with increasing CC size up to 
$R_{\rm pl}^{\rm CC}$ = 75 pc for $M^{\rm CC}$ = 1.0~$\times$ 10$^{6}$ and up to 100 pc 
for the more massive models. For larger values of $R_{\rm pl}^{\rm CC}$ the effective 
radii decrease rapidly (Figure \ref{figmassreff}). The turnover in the $r_{\rm eff}$ vs. 
$M_{\rm encl}$ space (Figures \ref{turnover} and \ref{obssims}) leads to degenerate states, 
as relatively compact CCs can produce a comparable merger object as a more massive CC 
having a significantly larger size. Despite the large range of CC sizes 
($R_{\rm pl}^{\rm CC}$ = 25 to 300 pc) of the models with initial CC masses of 
$M^{\rm CC}$ = 2.0~$\times$ 10$^{6}$, the effective radii of the merger objects are 
constrained to values between 13.2 and 25.6 pc.

In consequence, a range of initial conditions can form a merger object comparable to NGC\,2419 
preventing us to pinpoint the parameters of the original CC, which formed NGC\,2419. 
On the other hand, the larger the range of initial conditions that end up in a NGC\,2419-like object, 
the larger is the probability of creating a massive EC like NGC\,2419.

Due to the limited field of view of the Hubble Space Telescope, most extragalactic 
studies on GCs cover only (a part of) the optical disk of the respective galaxies. An outer 
halo object like NGC\,2419 would be found only by chance if it is projected onto the main 
body of a galaxy. Another reason for incompleteness is the difficulty of distinguishing ECs 
from background galaxies. Therefore, a number of surveys applied a size limit of about 10 pc 
to reduce the contamination of background galaxies. Thereby these surveys exclude all NGC\,2419-like
ECs from their GC catalogues. As massive galaxy-galaxy interactions are expected to have been more 
numerous in the past, a large number of NGC\,2419-like objects probably await their detection in the 
outer halos of various galaxies.

We conclude that NGC\,2419 can be well explained by the merged cluster complex scenario. 
Measurements of the proper motion of NGC\,2419 are indispensable to further study the proposed 
scenario and to potentially associate NGC\,2419 with one of the stellar streams in the 
outer Galactic halo.

\acknowledgments
We thank Manuel Metz for providing his \sbpp\ code and his for excellent support. 
We thank the anonymous referee for his helpful comments that helped us to improve the paper significantly.
The work of this paper was supported by DFG Grants KR\,1635/14-1 and KR\,1635/29-1.

\clearpage

\clearpage

\clearpage

\clearpage

\clearpage

\clearpage

\clearpage

\clearpage

\clearpage

\clearpage

\clearpage

\clearpage

\clearpage

\clearpage
\begin{deluxetable}{lrrrr}
\tablecolumns{11}
\tablewidth{0pc}
\tablecaption{Observational Parameters of NGC\,2419\label{tbl-1}}
\tablehead{
\colhead{} & \colhead{BC\tablenotemark{a}} &  \colhead{B\tablenotemark{b}} & 
\colhead{MvdM\tablenotemark{c}} &  \colhead{H\tablenotemark{d}}
}\startdata
Absolute V magnitude $M_{\rm V}$ [mag] & -9.28 & -9.4 & -9.31 & -9.58 \\
Total mass\tablenotemark{e} [10$^6$ M$_{\odot}$]& 0.90 & 1.01 & 0.93 & 1.19 \\
Central surface brightness $\mu_{\rm V}$ [mag arcsec$^{-2}$] & 19.61 & 19.55 & 19.44 & 19.83 \\
Core radius $r_{\rm c}$ [pc] & 7.8 & 7.8 & 7.9 & 8.57 \\
Effective radius $r_{\rm eff}$ [pc] & 19.2 & 23.5 & 19.9 & 17.88\\
Tidal radius $r_{\rm t}$ [pc] & 190 & 174 & 204 & 214\\
\enddata
\tablenotetext{a}{\cite{baumgardt}}
\tablenotetext{b}{\cite{bellazzini}}
\tablenotetext{c}{\cite{mvdm}}
\tablenotetext{d}{\cite{harris}}
\tablenotetext{e}{using the mass-to-light ratio of 2.05 from \cite{baumgardt}}
\end{deluxetable}

\begin{deluxetable}{lr}
\tablecolumns{11}
\tablewidth{0pc}
\tablecaption{Initial Cluster Complex and Star Cluster Parameters\label{tbl-inipar}}
\tablehead{
\colhead{} & \colhead{Parameter range}}
\startdata
Star Cluster (SC) & \\
\hline
Number of star cluster particles, $N_{\rm 0}^{\rm SC\,}$ & 100\,000\\
Initial SC mass, $M^{\rm SC\,}$ [10$^5$ M$_{\odot}$]& 0.5 -- 1.5 \\
Plummer radius of the SC, $R_{\rm pl}^{\rm SC\,}$ [pc]  & 4 \\
Cutoff radius of the SC, $R_{\rm cut}^{\rm SC\,}$ [pc]  & 20 \\
\hline
Cluster Complex (CC) & \\
\hline
Number of star clusters, $N_{\rm 0}^{\rm CC\,}$ & 20\\
Initial CC mass, $M^{\rm CC\,}$ [10$^6$ M$_{\odot}$]& 1.0 -- 3.0 \\
Plummer radius of the CC, $R_{\rm pl}^{\rm CC\,}$ [pc]  & 25 -- 300 \\
Cutoff radius of the CC, $R_{\rm cut}^{\rm CC\,}$ [pc]  & 100 -- 1200 \\
$\alpha$-parameter & 0.160 -- 0.013\\
\enddata
\end{deluxetable}

\begin{deluxetable}{lrrrrrrrr}
\tablecolumns{11}
\tablewidth{0pc}
\tablecaption{Parameters of the Merger Objects \label{tbl-2}}
\tablehead{
\colhead{Model\tablenotemark{a}} &  
\colhead{$N_{\rm M\,}$\tablenotemark{b}} &
\colhead{$M_{\rm encl\,}$\tablenotemark{c}} & \colhead{$M_{\rm V\,}$\tablenotemark{d}} &
\colhead{$r_{\rm h\,}$\tablenotemark{e}} & \colhead{$r_{\rm eff\,}$\tablenotemark{f}} &
\colhead{$r_{\rm c\,}$\tablenotemark{g}} &
\colhead{$\mu_{\rm V\,}$\tablenotemark{h}} & \colhead{$\sigma$\tablenotemark{i}}\\
\colhead{ } &\colhead{} &
\colhead{(10$^6$ M$_{\odot}$)} & \colhead{(mag)} &
\colhead{(pc)}  & \colhead{(pc)} & \colhead{(pc)} &
\colhead{(mag arcsec$^{-2}$)}  & \colhead{(km s$^{-1}$)} 
}
\startdata
M\_1\_1.0\_25   & 20 & 0.93 & --9.31 & 18.8 & 14.2 & 6.6 & 18.99 & 4.88\\
M\_1\_1.0\_50   & 20 & 0.85 & --9.21 & 25.8 & 19.2 & 5.7 & 19.36 & 4.13\\
M\_1\_1.0\_75   & 20 & 0.77 & --9.11 & 29.8 & 21.8 & 5.1 & 19.41 & 3.80\\
M\_1\_1.0\_100  & 19 & 0.64 & --8.91 & 24.9 & 19.1 & 5.5 & 19.51 & 3.59\\
M\_1\_1.0\_125  & 16 & 0.54 & --8.72 & 22.1 & 16.6 & 5.0 & 19.47 & 3.51\\
M\_1\_1.0\_150  & 14 & 0.46 & --8.55 & 19.6 & 14.4 & 4.5 & 19.39 & 3.47\\
\hline
M\_1\_1.5\_25   & 20 & 1.38 & --9.74 & 19.9 & 14.9 & 6.7 & 18.65 & 5.85\\
M\_1\_1.5\_50   & 20 & 1.29 & --9.67 & 26.6 & 20.0 & 6.2 & 19.04 & 4.99\\
M\_1\_1.5\_75   & 20 & 1.17 & --9.56 & 30.1 & 22.5 & 5.6 & 19.12 & 4.57\\
M\_1\_1.5\_100  & 19 & 1.05 & --9.44 & 31.8 & 23.5 & 4.9 & 19.08 & 4.34\\
M\_1\_1.5\_125  & 17 & 0.92 & --9.30 & 28.8 & 21.3 & 5.3 & 19.15 & 4.21\\
M\_1\_1.5\_150  & 15 & 0.80 & --9.15 & 25.1 & 18.8 & 4.8 & 19.09 & 4.10\\
\hline
M\_1\_2.0\_25   & 20 & 1.84 & --10.05& 21.1 & 15.9 & 7.2 & 18.47 & 6.55\\
M\_1\_2.0\_50   & 20 & 1.75 & --10.00& 28.4 & 21.2 & 6.4 & 18.83 & 5.65\\
M\_1\_2.0\_75   & 20 & 1.61 & --9.91 & 32.0 & 23.7 & 6.0 & 18.88 & 5.26\\
M\_1\_2.0\_100  & 19 & 1.48 & --9.82 & 35.2 & 25.6 & 5.3 & 18.87 & 4.99\\
M\_1\_2.0\_125  & 16 & 1.26 & --9.64 & 31.5 & 23.2 & 5.4 & 19.01 & 4.76\\
M\_1\_2.0\_150  & 18 & 1.17 & --9.56 & 28.9 & 21.2 & 5.1 & 18.81 & 4.81\\
M\_1\_2.0\_200  & 14 & 0.99 & --9.38 & 26.4 & 18.6 & 4.7 & 18.85 & 4.64\\
M\_1\_2.0\_300  & 10 & 0.62 & --8.87 & 17.6 & 13.2 & 4.8 & 19.03 & 4.18\\
\hline
M\_1\_3.0\_100  & 19 & 2.40 & --10.34& 39.6 & 28.8 & 5.3 & 18.45 & 6.08\\
\hline
M\_2\_1.5\_100  & 19 & 1.17 & --9.56 & 37.2 & 27.5 & 5.8 & 19.34 & 4.24\\
M\_3\_1.5\_100  & 19 & 1.07 & --9.46 & 31.9 & 23.7 & 5.3 & 19.21 & 4.33\\
M\_4\_1.5\_100  & 18 & 0.80 & --9.15 & 22.7 & 17.9 & 5.8 & 19.23 & 4.10\\
M\_5\_1.5\_100  & 19 & 0.87 & --9.24 & 31.3 & 24.5 & 5.4 & 19.35 & 3.84\\
M\_6\_1.5\_100  & 15 & 0.76 & --9.09 & 33.4 & 25.5 & 5.2 & 19.58 & 3.53\\
M\_7\_1.5\_200  & 19 & 0.93 & --9.31 & 30.1 & 21.9 & 5.1 & 19.12 & 4.22\\
\enddata
\tablenotetext{a}{Model\_\_Configuration\_\_$M^{\rm CC}$\_\_$R^{\rm CC}_{\rm pl}$.}
\tablenotetext{b}{Number of merged star clusters.}
\tablenotetext{c}{Enclosed mass of merger object within 800 pc.}
\tablenotetext{d}{Absolute V-magnitude of merger object.}
\tablenotetext{e}{Half-mass radius of merger object.}
\tablenotetext{f}{Effective radius, i.e. the projected half-mass radius of the merger object.}
\tablenotetext{g}{Core radius of merger object obtained from a King fit.}
\tablenotetext{h}{Central surface brightness $\mu_{V}$ of merger object obtained from a King fit. 
An extinction of $A_{\rm V}$ = 0.25 has been applied.}
\tablenotetext{i}{Velocity dispersion within a projected radius of 100 pc.}
\end{deluxetable}

\begin{deluxetable}{lrrrrrrrr}
\tablecolumns{11}
\tablewidth{0pc}
\tablecaption{$\beta$-values of our initial CC models\label{tbl-beta}}
\tablehead{
\colhead{$\beta$} & \colhead{$25$ pc\tablenotemark{a}} &  \colhead{$50$ pc} &
\colhead{$75$ pc} & \colhead{$100$ pc} & \colhead{$125$ pc} & \colhead{$150$ pc} & \colhead{$200$ pc}&
\colhead{$300$ pc}}
\startdata
$1.0 \times 10^6$ M$_{\odot}$\tablenotemark{b} & 0.38 & 0.77 & 1.19 & 1.63 & 2.14 & 2.73 & & \\
$1.5 \times 10^6$ M$_{\odot}$                  & 0.33 & 0.68 & 1.03 & 1.41 & 1.83 & 2.29 & & \\
$2.0 \times 10^6$ M$_{\odot}$                  & 0.31 & 0.62 & 0.94 & 1.28 & 1.64 & 2.04 & 3.02 & 7.27 \\
$3.0 \times 10^6$ M$_{\odot}$                  &  &  &  & 1.11 & & & &  \\
\enddata
\tablenotetext{a}{Plummer radius $R_{\rm pl}^{\rm CC\,}$ of the CC.}
\tablenotetext{b}{Initial CC mass $M^{\rm CC\,}$.}
\end{deluxetable}


\begin{thebibliography}{}
\bibitem[Aarseth et al.(1974)]{aarseth} Aarseth, S.J., Henon, M., \& Wielen, R. 1974, \aap, 37, 183
\bibitem[Bastian et al.(2006)]{bastian06} Bastian, N., Emsellem, E., Kissler-Patig, M., \& Maraston, C. 2006, \aap, 445, 471
\bibitem[Bastian et al.(2009)]{bastian09} Bastian, N., Trancho, G., Konstantopoulos, I.S. Miller, B.W. 2009, \apj, 701, 607
\bibitem[Baumgardt et al.(2009)]{baumgardt} Baumgardt, H., C\^ot\'e, P., Hilker, M., Rejkuba, M., Mieske, S., Djorgovski, S. G., \& Stetson, P.
2009, \mnras, 396, 2051
\bibitem[Bellazzini(2007)]{bellazzini} Bellazzini, M. 2007, \aap, 473, 171
\bibitem[Belokurov et al.(2006)]{belekurov} Belokurov, V. et al. 2006, \apj, 642, L137
\bibitem[Bien et al.(1991)]{bien} Bien, R., Fuchs, B., Wielen, R. 1991, in CP90 Europhysics Conference on Computational Physics,
ed. Tenner (Singapore, World Scientific), 3
\bibitem[Brodie \& Strader(2006)]{brodie06} Brodie, J. P., \& Strader, J. 2006, \araa, 44, 193
\bibitem[Br\"uns et al.(2009)]{bruens09} Br\"uns, R. C., Kroupa, P., \& Fellhauer, M. 2009, \apj, 702, 1268
\bibitem[Br\"uns et al.(2011)]{bruens11} Br\"uns, R. C., Kroupa, P., Fellhauer, M., Metz, M., \& Assmann, P.
2011, \aap, submitted
\bibitem[Cao \& Wu(2007)]{cao} Cao, C., \& Wu, H. 2007, \aj, 133, 1710
\bibitem[Casetti-Dinescu et al.(2009)]{casetti} Casetti-Dinescu, D.I., Girard, T.M., Majewski, S.R., Vivas, A.K., Wilhelm, R., Carlin, J. L., Beers, T. C., \& van Altena, W. F.
2009, \apj, 701, L29
\bibitem[Catelan (2009)]{catelan09} Catelan, M. 2009, \apss, 320, 261
\bibitem[Chandar et al.(2004)]{chandar04} Chandar, R., Whitmore, B, \& Lee, M. G. 2004, \apj, 611, 220
\bibitem[Chou et al.(2007)]{chou07} Chou, M.-Y. et al. 2007, \apj, 670, 346
\bibitem[Da Costa et al.(2009)]{dacosta} Da Costa, G. S., Grebel, E. K., Jerjen, H., Rejkuba, M., \& Sharina, M. E.
2009, \aj, 137, 4361
\bibitem[de Grijs et al.(2003)]{de_grijs03} de Grijs, R., Anders, P., Bastian, N., Lynds, R., Lamers, H. J. G. L. M., \& O'Neil, E. J.
2003, \mnras, 343, 1285
\bibitem[Drinkwater et al.(2000)]{drinkwater00} Drinkwater, M. J., Jones, J. B., Gregg, M. D.,
Phillipps, S. 2000, \pasa, 17, 227
\bibitem[Duffau et al.(2006)]{duffau} Duffau, S., Zinn, R. Vivas, A. K., Carraro, G., M\'endez, R. A., Winnick, R., \& Gallart, C.
 2006, \apj, 636, L97
\bibitem[Evstigneeva et al.(2007)]{evstigneeva07} Evstigneeva, E. A., Gregg, M. D., Drinkwater, M. J., Hilker, M. 2007, \aj, 133, 1722
\bibitem[Fellhauer et al.(2000)]{fell00} Fellhauer, M., Kroupa, P., Baumgardt, H., Bien, R., Boily, C. M., Spurzem, R., \& Wassmer, N.
2000, \na, 5, 305
\bibitem[Fellhauer et al.(2002)]{fell02a} Fellhauer, M., Baumgardt, H., Kroupa, P., \& Spurzem, R. 2002, CeMDA, 82, 113
\bibitem[Fellhauer \& Kroupa(2002a)]{fellhauer02a} Fellhauer, M., \& Kroupa, P. 2002a, \mnras, 330, 642
\bibitem[Fellhauer \& Kroupa(2002b)]{fellhauer02b} Fellhauer, M., \& Kroupa, P. 2002b, \aj, 124, 2006
\bibitem[Fellhauer \& Kroupa(2005)]{fellhauer05} Fellhauer, M., \& Kroupa, P. 2005, \mnras, 359, 223
\bibitem[Fellhauer et al.(2006)]{fellhauer06} Fellhauer, M., Kroupa, P., \& Evans, N. W. 2006, \mnras, 372, 338
\bibitem[Gallagher et al.(2001)]{gallagher01} Gallagher, S. C., Charlton, J. C., Hunsberger, S. D., Zaritsky, D., \& Whitmore, B. C.
 2001, \aj, 122, 163
\bibitem[Georgiev et al.(2009)]{georgiev} Georgiev, I.Y., Puzia, T.H., Hilker, M., Goudfrooij, P. 2009, \mnras, 392, 879 
\bibitem[Grillmair(2006)]{grillmair} Grillmair, C. J. 2006, \apj, 645, L37
\bibitem[Harris(1996)]{harris} Harris, W.E. 1996, \aj, 112, 1487 
\bibitem[Harris et al.(2002)]{harris02} Harris, W.E., Harris, G.L.H., Holland, S.T., \& McLaughlin, D.E. 2002, \aj, 124, 1435
\bibitem[Harris(2009)]{harris09} Harris, W.E. 2009, \apj, 699, 254
\bibitem[Ha\c{s}egan et al.(2005)]{hasegan} Ha\c{s}egan, M. et al. 2005, \apj, 627, 203
\bibitem[Hernquist(1990)]{hern1990} Hernquist, L. 1990, \apj, 356, 359
\bibitem[Hilker et al.(1999)]{hilker99} Hilker, M., Infante, L., Vieira, G., Kissler-Patig, M., \& Richtler, T.
  1999, \aaps, 134, 75
\bibitem[Homeier et al.(2002)]{homeier} Homeier, N., Gallagher, J. S. III, \& Pasquali, A. 2002, \aap, 391, 857
\bibitem[Huxor et al.(2004)]{huxor04} Huxor, A., Tanvir, N. R., Irwin, M., Ferguson, A., Ibata, R., Lewis, G., \& Bridges, T.
2004, ASPC, 327, 118
\bibitem[Huxor et al.(2008)]{huxor08} Huxor, A. P., Tanvir, N. R., Ferguson, A. M. N., Irwin, M. J., Ibata, R., Bridges, T., \& Lewis, G. F.
2008, \mnras, 385, 1989
\bibitem[Huxor et al.(2009)]{huxor09} Huxor, A., Ferguson, A.M.N., Barker, M.K., Tanvir, N.R., Irwin, M.J., Chapman, S.C., Ibata, R., Lewis, G. 2009, \apj, 698, 77
\bibitem[Jordan et al.(2005)]{jordan05} Jordan, A., Côté, P., Blakeslee, J.P., Ferrarese, L., McLaughlin, D.E., Mei, S., Peng, E.W., Tonry, J.L., Merritt, D., Milosavljevic, M., Sarazin, C.L., Sivakoff, G.R., West, M.J. 2005, \apj, 634, 1002
\bibitem[King(1962)]{king62} King, I. 1962, \aj, 67, 471
\bibitem[Konstantopoulos et al.(2009)]{konstantopoulos} Konstantopoulos, I. S., Bastian, N., Smith, L. J., Trancho, G., Westmoquette, M. S., \& Gallagher, J. S.
2009, \apj, 701, 1015
\bibitem[Kroupa(1998)]{krou98} Kroupa, P. 1998, \mnras, 300, 200
\bibitem[Kroupa(2008)]{krou08} Kroupa, P. 2008, in The Cambridge N-Body Lectures, Lecture Notes in Physics, Vol. 760, eds. S. Aarseth, C. Tout, \& R. Mardling
(Berlin: Springer Verlag), 181
\bibitem[Larsen et al.(2002)]{larsen02} Larsen, S. S., Efremov, Y. N., Elmegreen, B. G., Alfaro, E. J., Battinelli, P., Hodge, P. W., \& Richtler, T.
2002, \apj, 567, 896
\bibitem[Mackey \& Gilmore(2004)]{mackey04} Mackey, A. D., \& Gilmore, G. F 2004, \mnras, 352, 153
\bibitem[Mackey et al.(2006)]{mackey06} Mackey, A. D. et al. 2006, \apj, 653, L105
\bibitem[Masters et al.(2010)]{masters10} Masters, K.L., Jordan, A., Côté, P., Ferrarese, L., Blakeslee, J.P., Infante, L., Peng, E.W., Mei, S., West, M. 2010, \apj, 715, 1419
\bibitem[McLaughlin \& van der Marel(2005)]{mvdm} McLaughlin, D. E., \& van der Marel, R. P. 2005, \apjs, 161, 304
\bibitem[Mengel et al.(2008)]{mengel08} Mengel, S., Lehnert, M. D., Thatte, N. A., Vacca, W. D., Whitmore, B., \& Chandar, R.
2008, \aap, 489, 1091
\bibitem[Metz(2008)]{metz} Metz,~M. 2008, PhD thesis, Universit\"at Bonn
\bibitem[Mieske et al.(2008)]{mieske08} Mieske, S. et al. 2008, \aap, 487, 921
\bibitem[Miyamoto \& Nagai(1975)]{miya1975} Miyamoto, M., \& Nagai, R. 1975, \pasj, 27, 533
\bibitem[Newberg et al.(2003)]{newberg03} Newberg, H.J. et al. 2003, \apj, 596, L191
\bibitem[Newberg et al.(2009)]{newberg09} Newberg, H.J., Yanny, B., \& Willett, B.A. 2009, \apj, 700, L61
\bibitem[Newberg et al.(2010)]{newberg10} Newberg, H.J., Willett, B.A., Yanny, B., \& Xu, Y. 2010, \apj, 711, 32
\bibitem[Pellerin et al.(2010)]{pellerin} Pellerin, A., Meurer, G. R., Bekki, K., Elmegreen, D. M., Wong, O. I., \& Knezek, P. M. 2010, \aj, 139, 1369
\bibitem[Plummer(1911)]{plum1911} Plummer, H. C. 1911, \mnras, 71, 460
\bibitem[Richtler et al.(2005)]{richtler} Richtler, T., Dirsch, B., Larsen, S., Hilker, M., Infante, L. 2005, \aap, 439, 533
\bibitem[Ripepi et al.(2007)]{ripepi} Ripepi, V. et al. 2007 \apj, 667, L61
\bibitem[Salaris \& Weiss(2002)]{salaris} Salaris, M., \& Weiss, A. 2002, \aap, 388, 492
\bibitem[Stonkut\.{e} et al.(2008)]{stonkute} Stonkut\.{e}, R. et al. 2008, \aj, 135, 1482
\bibitem[van den Bergh(1995)]{vandenbergh95} van den Bergh, S. 1995, \aj, 110, 1171
\bibitem[van den Bergh \& Mackey(2004)]{vandenbergh04} van den Bergh, S., \& Mackey, A. D. 2004, \mnras, 354, 713
\bibitem[Whitmore \& Schweizer(1995)]{whitmore95} Whitmore, B. C., \& Schweizer, F. 1995, \aj, 109, 960
\bibitem[Whitmore et al.(1999)]{whitmore99} Whitmore, B. C., Zhang, Q., Leitherer, C., Fall, S. M., Schweizer, F., \& Miller, B. W.
 1999, \aj, 118, 1551
\end{thebibliography}
\end{document}